\documentclass[12pt]{article}

\usepackage{newtxtext,newtxmath}
\usepackage{graphicx, pdflscape}

\usepackage[letterpaper,margin=1in]{geometry}

\renewenvironment{abstract}
	{\quotation}
	{\endquotation}

\date{\today
    }

\makeatletter
\renewcommand{\fnum@figure}{\textbf{Figure \thefigure}}
\renewcommand{\fnum@table}{\textbf{Table \thetable}}
\makeatother

\usepackage{scicite}

\usepackage{color}
\usepackage{siunitx}
\usepackage{mathtools}
\usepackage{longtable}
\usepackage{comment}
\usepackage{xspace}

\def\headingsm#1{\noindent\underline{#1}\\}
\newcommand{\knit}{\(\mathrm{K}\)\xspace}
\newcommand{\purl}{\(\mathrm{P}\)\xspace}

\def\scititle{
    Chiral Analogues of Knit Stitches Designed Using Chiral Topology
}
\title{\bfseries \boldmath \scititle}

\def\sciauthor{
	Shunsuke~Takano$^{1,2,\ast}$,
	Yusuke~Kochi$^{1,3}$,
    Ken’ichi~Yoshida$^{1}$,
    Elisabetta~A.~Matsumoto$^{1, 4}$,
    \and
	Yuka~Kotorii$^{1,5,6}$,
    Toru~Asahi$^{7,8,9}$,
    Katsuya~Inoue$^{1,5,10,\dagger}$
}
\def\sciaffiliation{
	\small$^{1}$
    International Institute for Sustainability with Knotted Chiral Meta Matter (WPI-SKCM$^{\text{2}}$), 
    \and \small Hiroshima University, Higashi-Hiroshima, 739-8526, Japan.
    \and
	\small$^{2}$
    Graduate School of Advanced Science and Engineering, Waseda University, 
    \and \small TWIns, 2-2 Wakamatsu-cho, Shinjuku, Tokyo, 162-8480, Japan.
    \and
    \small$^{3}$
    Faculty of Science, Hiroshima University, Higashi-Hiroshima, 739-8526, Japan.
    \and
    \small$^{4}$
    School of Physics, Georgia Institute of Technology, Atlanta, GA, 30332, USA.
    \and
    \small$^{5}$
    Graduate School of Advanced Science and Engineering, 
    \and \small Hiroshima University, Higashi-Hiroshima, 739-8526, Japan.
    \and
    \small$^{6}$
    RIKEN, Interdisciplinary Theoretical and Mathematical Sciences Program (iTHEMS), 
    \and \small 2-1 Hirosawa, Wako, Saitama 351-0198, Japan.
    \and
    \small$^{7}$
    Faculty of Science and Engineering, Waseda University, 3-4-1 Okubo, Shinjuku, Tokyo, 169-8555, Japan
    \and
    \small$^{8}$
    Comprehensive Research Organization, Waseda University, 
    \and \small TWIns, 2-2 Wakamatsu-cho, Shinjuku, Tokyo, 162-8480, Japan.
    \and
    \small$^{9}$
    Research Organization for Nano \& Life Innovation, Waseda University; Tokyo, 162-0041, Japan.
    \and
    \small$^{10}$
    Chirality Research Center (CResCent), Hiroshima University, Higashi-Hiroshima, 739-8530, Japan.
}
\def\sciemail{
	\small$^\ast$
    shunsuke.t-8395@akane.wasda.jp
    \and
    \small$^\dagger$
    kxi@hiroshima-u.ac.jp
}
\author{
    \sciauthor
    \and
    \sciaffiliation
    \and
    \sciemail
}

\begin{document} 

% Insert the title and author list
\maketitle

% Abstract, in bold
% There are strict length limits, and not all formats have abstracts.
% Consult the journal instructions to authors for details.
% Do not cite any references in the abstract.
\begin{abstract} \bfseries \boldmath
    Fabrics are flexible thin structures made of  entangled yarn or fibers, yet the topological bases of their mechanics remain poorly understood. For weft knitted fabrics, we describe how the entanglement of adjacent stitches contributes to the flexibility of the fabric. Interpreting heterogeneous stitch pairs as domain boundaries reveals that the step between pairs of neighboring stitches is responsible for direction-specific flexibility. In typical knitted fabrics, anisotropic flexibility can be attributed to latticed domain boundaries. The intersections between domain boundaries result in point defects that induce frustration that resembles the impossible \textit{Penrose stairs}. We identify these by a chiral characteristic, defined summing the ascending or descending steps in a cycle surrounding the defect. Remarkably, seed fabric, a knit with high flexibility in both course and wale directions, is characterized as a racemic crystal of these chiral point defects.
    
    %\textcolor{cyan}{How to convert this manuscript: .tex $\to$ .pdf $\to$ .docx}
% Start with one or two sentences of background This is a simple template to prepare papers in \LaTeX\ for the \textit{Science}-family journals. Abstracts start with one or two sentences of background, which should be comprehensible to any scientist.
% Then summarise the results of your observations, experiments, simulations etc. The following text should outline the main results of the research. Simple mathematical expressions can be included e.g. $a^2+b^2=c^2$.
% End with a statement of your main conclusions The final sentence of the abstract should state the main conclusions and implications.
\end{abstract}

% The first paragraph of any Science paper does NOT have a heading
% Nor is it indented
\noindent
    Two-periodic textiles are tangles that live on a \emph{thickened} or \emph{quasi-two-dimensional} rectangular lattice. The thickening of the lattice resolves the entanglements of the filaments in the textile to over- or under-crossings. The symmetries of quasi-two-dimensional systems with in-plane periodicity are classified using layer groups, of which there are 80 distinct types~\cite{Herrmann1929zkrist, Weber1929zkristallogr, IT_E2006, Markande2020bridges, Diamantis2024Symmetry, DeLasPenas2024acta}. 
    Their networks are topologically protected by the entanglement of the yarn, and the fabric can bend and stretch without being disentangled. 
    The yarn and the way it is organized into entangled networks regulate the functionalities of different fabrics, such as bending rigidity, tear resistance, direction-specific elasticity, and density. 
    Many techniques have been devised over millennia for various applications, including garments, basketry, and fishing nets. 
    Among them, woven fabrics, which are predominantly inextensible, are assembled from crossed warp and weft threads, including the simplest plain weave and the slightly stretchable twill weave, which are adopted to canvas and denim, respectively. 
    Prehistoric artifacts indicate that the robust herringbone weave, called \textit{ajiro}, and the hexagonal weave, known as \textit{kagome}, have been frequently used for basketry~\cite{Sannai1998tower, Noshiro2019archaeol}.
    Knitted fabrics are elastic textiles formed with loops of yarn, called \emph{stitches}, as the primary building blocks. 
    The most common structures used in weft knitting include uniform stockinette, often used for cut-and-sew garments, and rib, often used for sweater cuffs due to their high flexibility in the weft direction~\cite{Pfister1945dura}. 
    The topology of a material affects its properties. 

    The topology of a material can also affect its properties. A periodic network of yarn in a fabric can be viewed as a knot or link in the thickened torus~\cite{Grishanov2009textres, Markande2020bridges, Diamantis2024Symmetry}, and the network of entangled yarn is analyzed with knot theory~\cite{Kauffman2006formal, Purcell2020Hyperbolic}. Knot theory concerns the topology of closed loops, and two knots are \emph{isotopic} if they smoothly transform from one to the other. 
    For links, invariants such as the linking number, Kauffman polynomials, and Jones polynomials are useful to distinguishing them~\cite{Kauffman1990transamermathsoc}. 
    The simplest non-trivial knot, the trefoil knot $3_1$, is chiral, namely, its Kauffman polynomial is different from its mirror image and, thus, is not isotopic to its mirror image.
    Chirality can induce distinct material properties, such as  determines the mechanical stability of knotted ropes~\cite{Patil2020sci}. In plied ropes, a \textit{lang lay}---defined by strands twisted in the same direction as they wind around the core, resulting in a homochiral arrangement---is more flexible and exhibits a $15\text{--}20\%$ longer service life under bending than a \textit{regular lay}, which has strands twisting in the opposite direction as they wind around the core, resulting in a heterochiral structure tht is subject to higher internal pressure~\cite{WIREROPEUSERSMANUAL2nd1981}. Since chirality imparts flexibility to knots and ropes, we predict it will similarly enhance the flexibility of fabrics.

    The induced three-dimensional structure greatly increases the extensibility of the materials.
    Polymer blobs~\cite{deGennes1979polymer}, the Miura fold~\cite{Miura2020form}, and certain single-celled protists~\cite{Flaum2024sci} exhibit unusually high deformability that arises from folded chains, folded sheets, and folded bundles of microtubules, respectively.
    Knitted fabrics are generally more extensible than woven textiles, as knits have a richer abundance of meandering yarn paths. 
    Under tension, these meandering paths accommodate elongation primarily through the bending and straightening of the yarn centerline, rather than through intrinsic stretching of the yarn itself. 
    This geometric principle is effective in enhancing the extensibility of different knitted fabrics. 
    Stockinette, rib, garter, and seed are typical weft-knitted fabrics, each formed simply by varying the arrangement of \textit{knit} (\knit) and \textit{purl} (\purl) stitches. 
    Even when made from the same yarn, the arrangements differentiate fabrics' elastic responses~\cite{Singal2024natcommun}: rib and garter have more meandering yarn, making them flexible more than stockinette. Stockinette is composed entirely of \knit stitches, resulting in relatively little yarn meandering. 
    However, when fragments of stockinette and reverse stockinette are joined together, the boundaries between \knit- and \purl-domains induce out-of-plane zig-zags, which force the yarn to meander, thereby imparting increased extensibility at the domain boundary~\cite{Singal2024natcommun}. Rib and garter inherently contain multiple domain boundaries that promote greater yarn meandering and contribute to their greater flexibility. In traditional knitted stitches, however, the minimum spatial interval between successive over–under crossings limits the extent of yarn meandering within a single stitch. Here, we introduce chiral knit analogues by reducing this interval, thereby increasing the amplitude of out-of-plane zig-zags. This enhances the overall meandering of the yarn paths, which in turn further increases the elastic response of the fabrics.

\section*{\label{sec:DomainBoundary}Domain boundaries and biaxial extensibility} %570 words
\begin{figure}[h]
        \centering
        \includegraphics{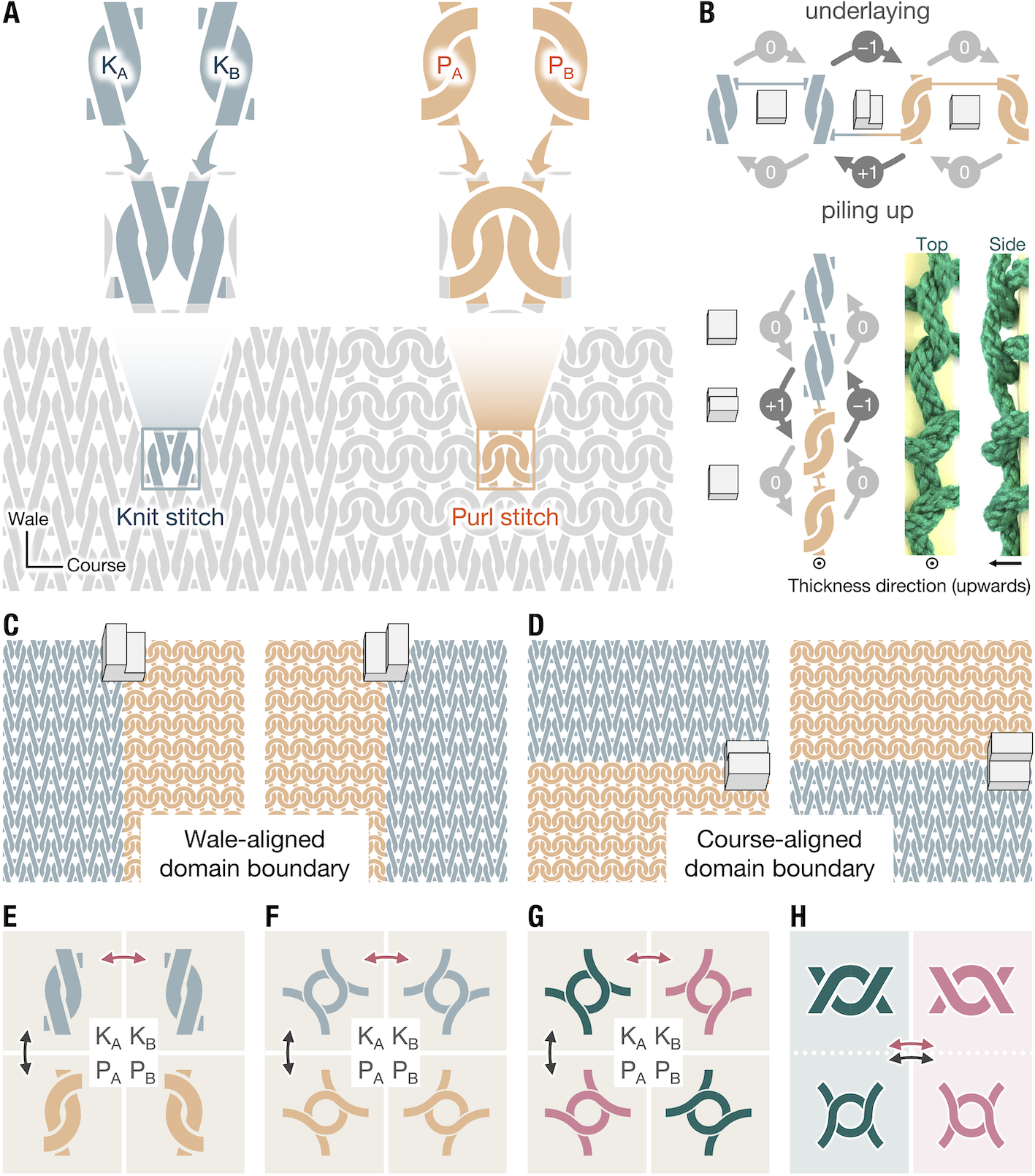}
        \caption{
            \label{fig:Stitch}
            \textbf{
                Knots comprising knitted fabrics.
            }
            (\textbf{A}) Diagram of a knitted fabric and two stitches, \knit and \purl, that are decomposed into four units, $\mathrm{K_A}$, $\mathrm{K_B}$, $\mathrm{P_A}$ and $\mathrm{P_B}$. 
            Diagram is a regular projection that depicts the overcrossing and undercrossing of yarn.
            (\textbf{B}) Relative out-of-plane positioning of units.
            (\textbf{C} and \textbf{D}) Domain boundaries as a step emerging at boundary of \knit- and \purl-domains. 
            (\textbf{E}--\textbf{H}) Symmetry of the units. 
            The black and pink arrows represent reflection on a mirror in and normal to the fabric plane, respectively.
            (F) shows another representation of the units, which are smoothly transformed from (E) while maintaining the crossings of the yarn. 
            (G) distinguishes the handedness of the units by the green and red coloring. 
            Pairs of the units in diagonal panels are equivalent by $C_2$ about the normal of the fabric plane. 
            (H) shows that green and red units are mirror images of each other.
            %\textcolor{magenta}{$120\times135.8\,\si{mm^2}\sim435$ words.}
            }%152 words
    \end{figure}

    Knitted fabrics can be described as a rectangular lattice of knots, where each lattice site is occupied by either a \knit or \purl stitch (Fig.~\ref{fig:Stitch}A)~\cite{Singal2024natcommun}. These stitches have a mirror plane along the wale direction. 
    Using mirror symmetry, each stitch can be divided into smaller asymmetric \textit{units}: the \knit stitch into $\mathrm{K_A}$ and its mirror image, $\mathrm{K_B}$, and the \purl stitch into $\mathrm{P_A}$ and its mirror image $\mathrm{P_B}$ (Fig.~\ref{fig:Stitch}A). 
    These four units are invariant under $C_2$ rotation.\footnote{Note that in a knitted textile, the units $\mathrm{K_A}$ and $\mathrm{K_B}$ or $\mathrm{P_A}$ and $\mathrm{P_B}$ must occur in pairs, as basic stitches are formed by pulling a loop of yarn through an existing loop in the fabric.} We begin by using these asymmetric units to characterize the ways in which a single stitch can be joined to its neighbors. 
    Here, we use the term \textit{adjacent} to refer to nearest neighbors in the course or wale direction\footnote{The course and wale directions correspond to the in-plane horizontal and vertical directions, respectively.
    In hand knitting, yarn progresses along the course, and loops are pulled and extend along the wale.} and does not include diagonal neighbors.
    We have characterized the relative out-of-plane positioning between a unit and its adjacent unit, see Supplementary Materials~\cite{supplmater}. 
    A unit has four yarn segments that connect it to  adjacent units. 
    The unit $\mathrm{K_\bullet}$ \footnote{The bullet symbol $\bullet$ denotes an unspecified index, with possible values $\mathrm{A}$ and $\mathrm{B}$.} consists of two segments that have two crossings. The four ends of the yarn segments join the unit to its nearest neighbors. The two ends that join the unit to its neighbors in the wale (or course) direction come from and over-crossing (or under-crossing) and lie above (or below) the segments of yarn connecting the unit to its neighbors in the course (or wale direction).
    By comparing the level of the crossings at either side of a yarn segment that joins neighboring units, we can develop a height ordering between neighboring units. 
    For example, consider neighboring units  $\mathrm{P_A}$ (left) and $\mathrm{K_B}$ (right) in the course direction. The yarn segment joining them  leaves $\mathrm{P_A}$ as an over-crossing and enters $\mathrm{K_B}$ as an under-crossing. This implies that the $\mathrm{K_B}$ unit is situated ``above'' the join and $\mathrm{P_A}$ is situated ``below it''.
    We describe this configuration  as $\mathrm{K_B}$ being ``one step above'' $\mathrm{P_A}$. Likewise, when $\mathrm{K_A}$ (left) and $\mathrm{K_B}$ (right) are neighboring in the course direction, they are both situated ``above'' the yarn segment connecting them (Fig,~\ref{fig:Stitch}B), and thus we consider $\mathrm{K_\bullet}$s to be on the same level as one another.

    Uniform weft knitted fabrics are known as stockinette and reverse stockinette, which correspond to monodomains formed by the \knit and \purl stitches, respectively (Fig.~\ref{fig:Fabrics}). 
    Since every unit is adjacent to units of the same kind, all the stitches are on the same level without steps. 
    Note that stockinette and reverse stockinette are identical as they are interconvertible by $C_2$ rotation around the wale axis, and both crystallize in the same layer group $pbm2\, (\text{L24})$. 
        
    Joining a stockinette domain to a reverse stockinette domain results in a $\mathrm{K_\bullet}\mathrm{P_\bullet}$ motif at the domain boundary (Fig.~\ref{fig:Stitch}C and D). 
    When $\mathrm{K}$- and $\mathrm{P}$-domains are placed in the course direction, the domain boundary is parallel to the wale (i.e., longitudinal) direction, making the $\mathrm{K}$-domain one step higher than the $\mathrm{P}$-domain (Fig.~\ref{fig:Stitch}C). 
    Since a heteropair of units is softer than a homopair, the domain boundary is more extensible than the stockinette or reverse stockinette monodomains. 
    The yarn segment connecting $\mathrm{K_\bullet}$ and $\mathrm{P_\bullet}$ has odd symmetry, namely, it forms an angle with respect to the fabric plane~\cite{Singal2024natcommun}. 
    Under tensile stress in the course direction, the angle of the adjoining yarn segment decreases. This enables the fabric to extend with minimal deformation to the yarn. 
    In this regime, the strain deformation of the fabric arises from the bending of the yarn. Note that the bending modulus may be two orders of magnitude smaller than the Young's modulus that governs yarn extension~\cite{supplmater}. 
    Wale-aligned domain boundaries are aligned in rib (Fig.~\ref{fig:Fabrics}). 
    The traditional rib has alternating \knit and \purl stitches in the course direction and shows greater extensibility than stockinette, especially in the course direction~\cite{Singal2024natcommun}.
    The high extensibility of rib is attributed to the heteropairs at the domain boundaries. 
    We observe a similar enhancement to extensibility in domains that alternate in the wale direction.
    Garter has parallel domain boundaries arranged in the course direction (Fig.~\ref{fig:Fabrics}), which shows an odd $\mathrm{K_\bullet}\mathrm{P_\bullet}$ arrangement across the domain boundary (Fig.~\ref{fig:Stitch}D). 
    The yarn segments connecting the heteropairs improve the extensibility of the fabric in the wale direction. 

%%%%%%%%%%%%%%%%%%%%%%%%%%%%%%%%%%%%%%%%%%%%%%%%%%%%%%%%%%%%%%%%%%%
\section*{\label{sec:PointDefect}Frustrated out-of-plane positioning of units around point defect} %357 words
    Point defects appear at intersections of the course- and wale-aligned domain boundaries and interplay additional mechanical properties of the fabric. 
    We define the \emph{vortex number} $v$ to be the sum of the rise of each step, where the rise of a step up is $+1$ and the rise of a step down is $-1$,
    along a closed loop around a point defect, oriented in a counterclockwise sense. 
    The vortex number $v$ is independent of path choice, see Supplementary Material~\cite{supplmater}. 
    Figure~\ref{fig:Defects}B depicts a point defect arising from four domains arranged around it. It lies at the intersection between domain boundaries in the course direction and the wale direction.
    The vortex number of the point defect is equal to $+4$ as the square path undergoes four ascents in one cycle. 
    Figure~\ref{fig:Defects}C shows the mirror image of the defect under a reflection through the plane of the fabric. Its  vortex number now becomes $-4$. 
    Since vortex number is transformed as $v\to -v$ under reflection, the vortex number can characterize the chirality of a point defect.
    A spiral staircase also behaves like a chiral point defect~\cite{Barron2012chirality} -- as users traverse a cycle around the staircase, they either ascend or descend by one floor, depending on the relative alignment of the loop and the chirality of the stairs(Fig.~\ref{fig:Defects}D). 
    Although a spiral staircase always takes users to another floor, a round trip around a point defect in a fabric returns to the initial level, similar to the Penrose staircase (Fig.~\ref{fig:Defects}E), known as an impossible object~\cite{Penrose1958psy}. 
    This paradox around a point defect indicates frustration in the out-of-plane positioning of the units. The units are forced to twist around the point defect, like blades of a propeller. 
    Consequently, the yarn segments connecting the units around the point defect are less prone to tilting perpendicular to the plane of the fabric, and the frustration, thus suppresses the extensibility of the fabric.

\begin{figure}[h]
        \centering
        \includegraphics{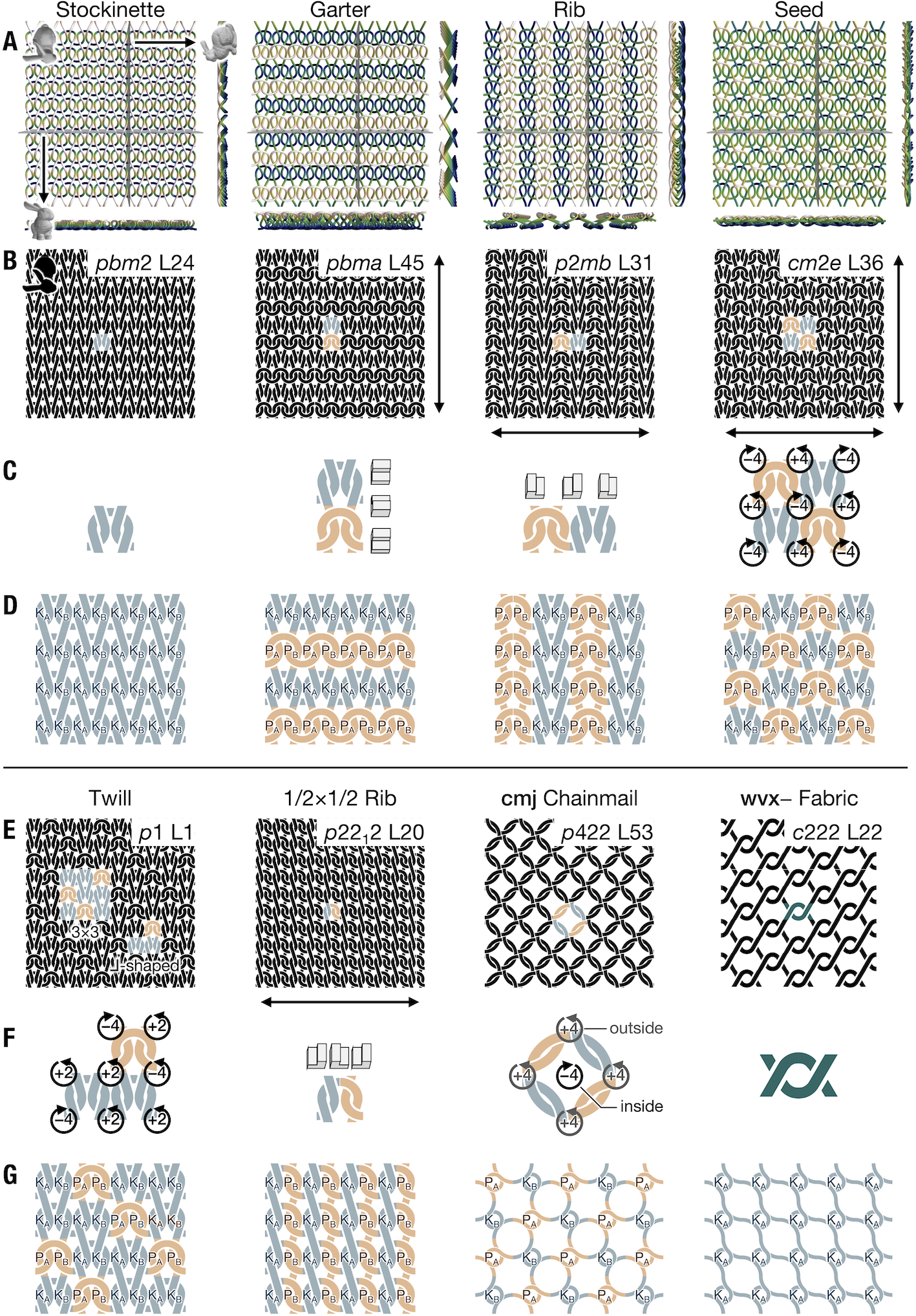}
        \caption{
            \label{fig:Fabrics}
            \textbf{Crystallographic analysis of weft knitted and similar fabrics.}
            (\textbf{A}) Top, front and side views of the four typical weft knitted fabrics. The Stanford bunny represents the angle of view, and gray planes show cut surfaces. 
            (\textbf{B} and \textbf{E}) Diagram of fabrics. 
            Layer group is labeled in the top right-hand corner, and repeating unit is highlighted in color. 
            The double arrows indicate the high elasticity direction.
            (\textbf{C} and \textbf{F}) Domain boundary and point defect in a repeating unit.
            (\textbf{D} and \textbf{G}) Arrangement of units. 
            $\mathrm{K_\bullet}$ and $\mathrm{P_\bullet}$ are distinguished by color and label on the illustrations.
            %\textcolor{magenta}{$120\times 172.8\,\si{mm^2}\sim 553$ words.}
            }%97 words
    \end{figure}

    In seed fabric, \knit and \purl stitches are arranged alternately as a checkerboard pattern, which generates domain boundaries  in both the course and wale directions (Fig.~\ref{fig:Fabrics}). 
    Point defects appear at the intersections between the course- and wale-aligned domain boundaries. 
    Point defects with $+4$ and $-4$ vortex numbers are distributed in a checkerboard pattern. 
    The domain boundaries make seed more flexible than uniform stockninette. 
    Given that point defects impart rigidity to fabrics, seed displays diminished extensibility compared with rib and garter in the course and wale directions, respectively~\cite{Singal2024natcommun}.

    \begin{figure}[h]
        \centering
        \includegraphics{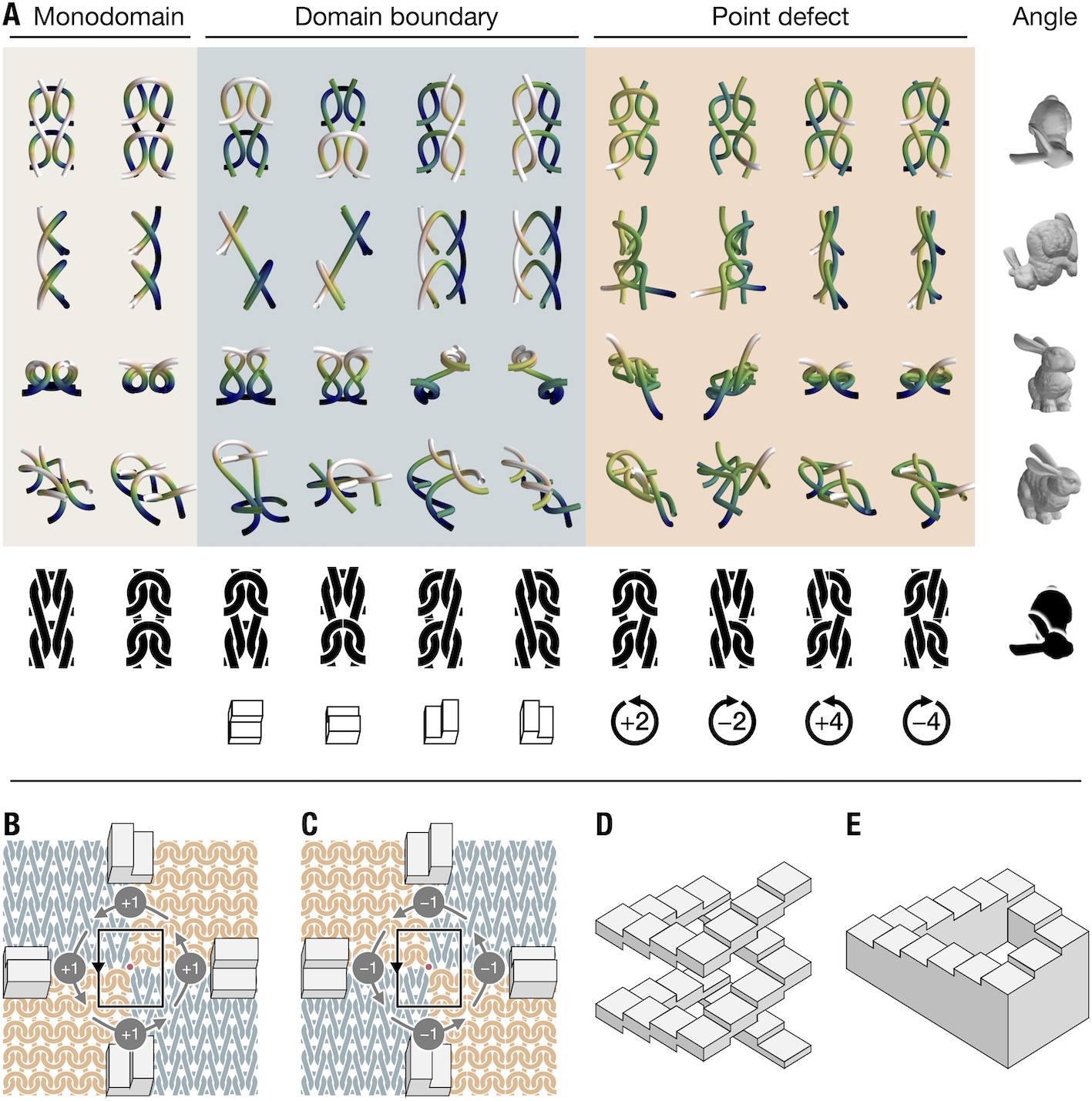}
        \caption{
            \label{fig:Defects}
            \textbf{
            Classification of defects emerging at the point where the four sites meet.
            }
            (\textbf{A}) Three-dimensional structure of defects. The Stanford Bunny represents the angle of view. The color of the yarn represents the position in the thickness direction. 
            (\textbf{B} and \textbf{C}) Point defects with vortex number $\pm4$. 
            A red point represents the center of the point defect in concern. 
            The vortex number is computed by summing steps along the black square line.
            (\textbf{D}) A spiral staircase.
            (\textbf{E}) The Penrose staircase.
            %\textcolor{magenta}{$120\times120.4\,\si{mm^2}\sim 385$ words.}
            }%79 words
    \end{figure}

%%%%%%%%%%%%%%%%%%%%%%%%%%%%%%%%%%%%%%%%%%%%%%%%%%%%%%%%%%%%%%%%%%%
%\section*{\label{sec:Weft}Chirality of weft}
%%%%%%%%%%%%%%%%%%%%%%%%%%%%%%%%%%%%%%%%%%%%%%%%%%%%%%%%%%%%%%%%%%%
\section*{\label{sec:Typical}Two interpretations of chirality in weft knitted fabrics}  %146 words
    Typical weft knitted fabrics are crystallographically achiral, namely they have mirror planes, glide planes or inversion centers (Fig.~\ref{fig:sup_layergroup}A--D). From the perspective of knot topology, the typical weft knitted fabrics are constructed from achiral constituents: \knit and \purl, which have mirror planes in the wale direction.
    The chirality of the fabrics can also be evaluated by the domain structures, especially point defects with finite vortex numbers. 
    Stockinette, rib, and garter have no point defects, and these fabrics are achiral. Seed has an array of point defects with vortex numbers $v=\pm 4$ and thus is locally chiral. 
    However, the point defects in seed always come in pairs with vortex number $v=+4$ and $v=-4$. Thus, if we look at a large enough sample of fabric, the net vortex number will cancel out. From the perspective of defect chirality, we say that seed fabric is racemic, since it is made up of equivalent numbers of chiral constituents. 
    
%=Weft/Chiral===========================================================================
\section*{\label{sec:Chiral} Enantiomeric knitting}  %714 words
    
    The four weft knitted fabrics we have previously discussed inherit mirror planes from their constituent stitches, since the repeated units are aligned parallel or perpendicular to the mirror plane in the wale direction. 
    While these fabrics are achiral, we can design chiral fabrics that do not have a global mirror symmetry. 
    Consider the ``twill'' fabric shown in Fig.~\ref{fig:Fabrics}F. We may choose a \textit{unit cell} with three stitches in both the course and the wale ($3\times3$ unit cell). The $3\times3$ unit cell contains six \knit and three \purl stitches, repeats along the course and wale at intervals of three stitches, and does not possess mirror planes, glide planes, or inversion centers. Consequently, twill inherits the same symmetry and is chiral. The twill crystallizes in triclinic $p1\, (\text{L1})$. The \textit{primitive unit cell} of the twill (Fig.~\ref{fig:sup_layergroup}E) is one-third the size of the $3\times3$ unit cell and is aligned obliquely to both the course and the wale. Through a continuous deformation of the primitive unit cell, another \reflectbox{\textsf{L}}-shaped primitive unit cell is obtained, which is composed of two \knit and one \purl stitches.
    This chiral twill fabric demonstrates that intraditional weft knitting composed of \knit and \purl stitches, the unit cell of a chiral fabric must have three or more stitches. 
    This is because \knit and \purl stitches are hetero-chiral pairs $\mathrm{K_A}$-$\mathrm{K_B}$ and $\mathrm{P_A}$-$\mathrm{P_B}$, respectively (Fig.~\ref{fig:Stitch}). 
    The chiral macroscopic structure of the twill fabric organized by achiral stitches is an analogue of chiral condensed matters formed by achiral constituents, such as the twist-bend nematic phase~\cite{Jakli2013liqcrystrev} and $\ensuremath{\alpha}$-quartz~\cite{Arago1811}. 
    The fabric may also be interpreted as a \textit{kryptoracemate}\footnote{A racemic structure crystallizing in a Sohncke type~\cite{supplmater} of crystallographic space group~\cite{Nespolo2021japplcrystallogr}.} when the units are treated as building blocks. Twill contains equal amounts of units with opposite handedness and is therefore a racemate. Although racemates usually adopt achiral crystal structures, twill exhibits a chiral one. Kryptoracemates seldom occur among chiral compounds~\cite{Rekis2020ActaCrystB}; however, we found that this is possible in the knittable twill fabric. 
    Remarkably, this twill fabric has two $v=+2$ point defects and one $v=-4$ point defect per primitive unit cell, and the total vortex numbers are canceled out. Also in the $3\times3$ unit cell, the total vortex numbers vanish.

\section*{\label{sec:ChiralAnalog} Chiral knit-inspired fabrics}
    What happens when we ignore the assumption that a stitch is made of a hetero-chiral pair $\mathrm{K_A}$-$\mathrm{K_B}$ or $\mathrm{P_A}$-$\mathrm{P_B}$.
    While such a fabric would be inconsistent with knitting, it can be manufactured using loop or n\r{a}lebinding (knotless netting). 
    Consider a fabric is made entirely of either $\mathrm{K_A}$ or $\mathrm{P_A}$ units. Such a fabric would not be superimposable on its mirror image. 
    We devise three strategies to design fabrics with unit cells made from enantiomeric constituents. 
    
    Our first strategy is to substitute one of the units in stockinette, for example $\mathrm{K_B}$ for $\mathrm{P_B}$.
    This results in the deracemization of the fabric (Fig.~\ref{fig:Fabrics}B), since the resulting units, $\mathrm{K_A}$ and $\mathrm{P_B}$, have the same handedness (Fig.~\ref{fig:Stitch}G). We call this deracemized fabric `$1/2\times 1/2$ rib', as its alternating arrangement of $\mathrm{K_\bullet}$ and $\mathrm{P_\bullet}$ in the course direction resembles a rib structure with a shortened interval. It belongs to $p22_12\, (\text{L20})$~\cite{supplmater}. 
        
    The second strategy is to modify the lattice. 
    Traditional knitting lives on a rectangular lattice, which, in the cases of knit and purl stitches may additionally be described as superposition of sublattices A and B aligned in the course direction, where A and B correspond to the index of the units (Fig.~\ref{fig:Fabrics}D). 
    We placed sites A and B in a checkerboard pattern with $\mathrm{P_A}$ and $\mathrm{K_B}$ at each site. 
    Following Liu \emph{et al.} 2018 ~\cite{Liu2018chemsocrev}, this fabric is known as \textbf{cmj} chainmail and is composed of interlocked rings (Fig.~\ref{fig:Fabrics}). 
    As shown in Fig.~\ref{fig:Fabrics}F, the point defects surrounding the four units are distinguished as being either inside (highlighted in black) or outside (gray) the ring. These point defects have vortex number of $+4$ and $-4$, respectively. 
    Note that the total vortex numbers do not canceled out in this fabric, and the chiral structure crystallizes in $p422\, (\text{L53})$. 
    Chainmail differs in curvature depending on the degree of chirality. In particular, \textbf{cmj} chainmail is non-planar~\cite{Klotz2024soft}, in contrast to the 4-in-1 pattern, which has historically been used for armor in Europe~\cite{Wijnhoven2021}. 

\begin{figure}[h]
        \centering
        \includegraphics{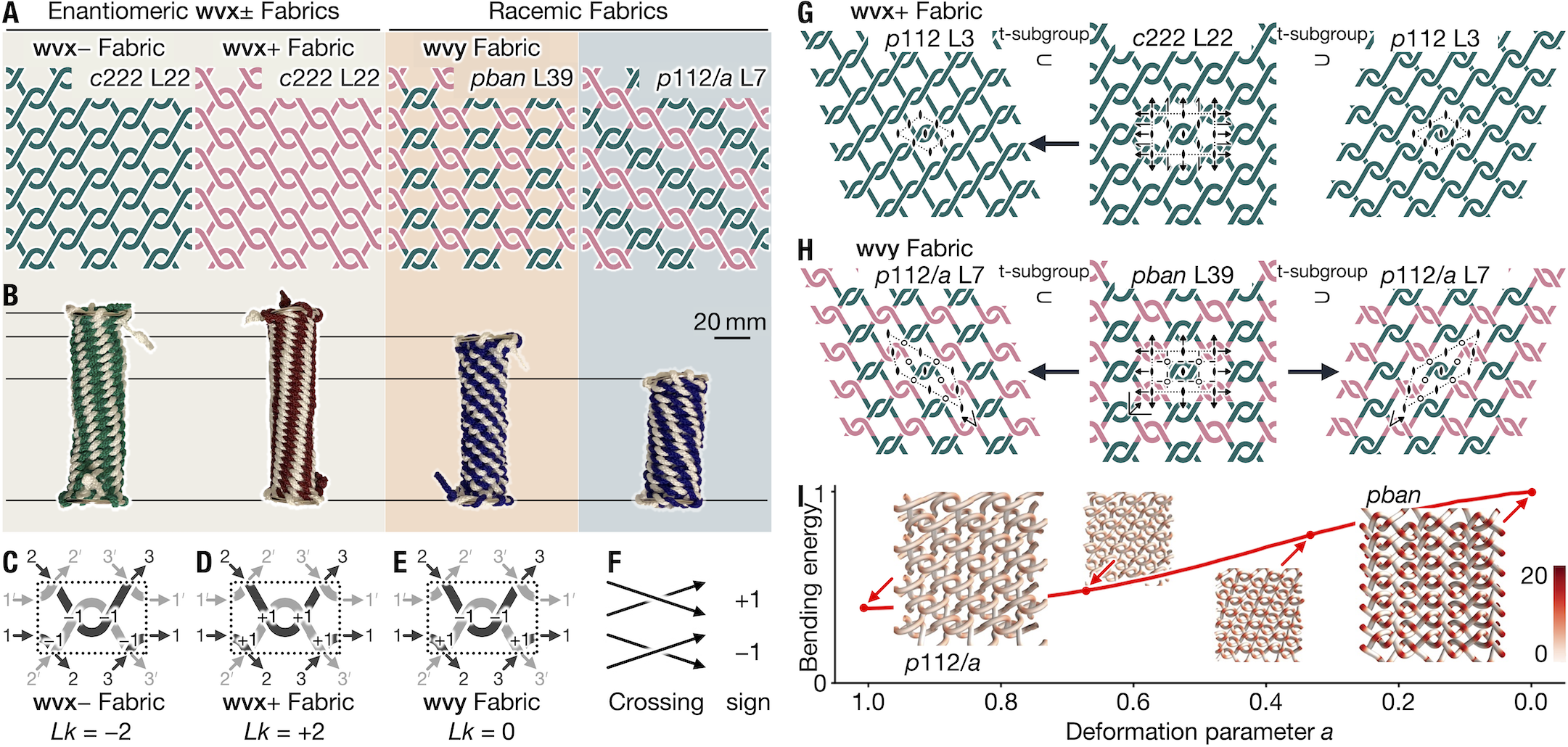}
        \caption{
            \label{fig:Chiralweft}
            \textbf{Enantiomeric knit-inspired fabrics and associated racemic fabric.}
            (\textbf{A}) Diagram and layer group of fabrics.
            (\textbf{B}) Tubular fabrics with the closed course direction. 
            (\textbf{C}--\textbf{E}) The linking number of the fabrics and (\textbf{F}) the sign of a crossing.  
            (\textbf{G} and \textbf{H}) Shear deformation and resulting changes in the symmetry of planar fabrics. 
            The arrows indicate the possible directions of spontaneous deformation.
            (\textbf{I}) Continuous shear deformation in \textbf{wvy} fabrics and change in the bending energy. The color of the yarn shows curvature. The bending energy is normalized by the length of yarn per unit cell, and by the energy in the initial state ($a=0$). 
            %\textcolor{magenta}{$183\times 87.6\,\si{mm^2}\sim 420$ words.}
            }%105 words
    \end{figure}

    The third strategy is to intertwine helices. 
    Interleaving helices of the same handedness form enantiomeric two-periodic fabrics, such as the \textbf{wvx}$\pm$ fabric described in~\cite{Liu2018chemsocrev}\footnote{We have added the $\pm$ modifier to the \textbf{wvx} notation to distinguish between fabrics made from left- and right-handed helices.} (Figs.~\ref{fig:Fabrics} and \ref{fig:Chiralweft}) belongs to $c222\, (\text{L22})$. The \textbf{wvx}$\pm$ fabric is composed of one type of unit ($\mathrm{K_A}$ in Figure~\ref{fig:Fabrics}B).
    The $2_1$ screw axes coincide with the helical axes of the spiraling yarn~\cite{supplmater}. 
    Helices may also form three-periodic weaves~\cite{Wetzel2024bridges}.
    Knot theory uses invariants, such as \emph{linking number} to distinguish pairs of links. If two links have different linking number, they are topologically distinct.\footnote{However, two links that have the same linking number are not guaranteed to be topologically equivalent.}  The linking number $Lk$  is defined as 
    \begin{equation}
        Lk(K_1, K_2)
        \coloneqq
        \frac{1}{2} \sum_{c\in D_1\cap D_2} \mathrm{sign}(c)
        ,
    \end{equation}
    where $K_1, K_2$ are two components of the textile link within a unit cell, $D_1, D_2$ are their corresponding components of the oriented planar link diagram. 
    The sign of each crossing $c$ is defined as shown in Fig.~\ref{fig:Chiralweft}F.\footnote{Note, the linking number $Lk$ is invariant under Reidemeister moves in $T^2 \times I$, which is the natural space for two-periodic textiles~\cite{supplmater}.} 
    The \textbf{wvx}$-$ fabric (shown in green in Fig.~\ref{fig:Chiralweft}B) has $Lk = -2$ (, and its mirror image, the \textbf{wvx}$+$ fabric (shown in red in Fig.~\ref{fig:Chiralweft}B), has $Lk = +2$ (Fig.~\ref{fig:Chiralweft}D). 
    The inversion of the sign under  reflection of the fabric indicates that \textbf{wvx}$\pm$ fabrics are chiral.
    Chiral \textbf{wvx}$\pm$ fabrics are topologically distinguished from weft knitted fabrics, which have $Lk=0$. 
    The intertwining of helices involves the use of yarn ends, a technique more akin to n\r{a}lebinding than to ordinary knitting, which relies solely on the manipulation of loops of yarn.
    Note that \textbf{wvx}$\pm$ fabrics show spontaneous shear deformations, which serves to more densely pack the yarn. Including the shear deformation, \textbf{wvx}$\pm$ fabrics crystallize in $p112\,(L3)$, the maximal t-subgroup with monoclinic/oblique lattice~\cite{supplmater} (Fig.~\ref{fig:Chiralweft}G).
    
%=Weft/Racemic===========================================================================
\section*{\label{sec:Racemic}Racemized fabrics}   %497 words
     Racemic fabrics must have units with different handedness in a one-to-one ratio. We introduce two methods to racemize the enantiomeric \textbf{wvx}$\pm$ fabrics.
    Our first fabric, a racemic version of \textbf{wvy} fabric (third from the left in Fig.~\ref{fig:Chiralweft}A), belonging to the centrosymmetric  $pban\, (\text{L39}),$ has a checkerboard pattern of $\mathrm{K_A}$ ($\mathrm{P_B}$) and $\mathrm{P_A}$ ($\mathrm{K_B}$). The linking number for the \textbf{wvy} fabric $Lk=0$ --- it is invariant under reflection in the plane of the fabric. 
    This racemic fabric has point defects with vortex numbers $\pm 4$, and the total vortex number is canceled out. The presence of these point defects suppress the elasticity. 
    The defects generate frustration that prohibits the unique determination of the levels of the individual units, making the fabric difficult to stretch. 
        
    The \textbf{wvy} fabric may show a spontaneous shear deformation which reduces the frustration. 
    Consider the unit cell formed from two strands of yarn, shown in Fig.~\ref{fig:Chiralweft}E. Let $\gamma_a: \mathbb{R}/\mathbb{Z} \times \mathbb{Z}_2 \to T^2 \times I$ be the centerline of a strand of yarn in \textbf{wvy} fabric, where $a \in [-1,1]$ is the deformation parameter, $s$ is the arclength parameter, and $l_n$ is the length of the $n^{\text{th}}$ strand of yarn within the unit cell, which satisfy $s/l_n \in \mathbb{R}/\mathbb{Z} \cong S^1$. 
    The deformation parameters $a=0$ and $a=\pm 1$ correspond to the initial flat and two sheared states, respectively. 
    A curve is characterized locally by curvature and torsion in the Frenet--Serret frame. Since increasing curvature costs  bending energy~\cite{Patil2020sci, Singal2024natcommun}, the initial state ($a=0$) with locally high curvature energetically unfavorable (Fig.~\ref{fig:Chiralweft}I). 
    When the fabric is sheared ($a = +1$), the units  $\mathrm{P_A}$ and $\mathrm{K_A}$ are offset slightly above and below the plane of the fabric, respectively. 
    This causes the yarn to zigzag in the thickness direction. 
    The curvature in zigzag yarn appears to be more uniformly distributed and smaller than the un-sheared state. Consequently, the decrease in total bending energy makes the shear deformation more energetically favorable. 
    Changing the deformation parameter corresponds to swapping the sense of the out-of-plane deformations of each unit---that is, for $a=-1,$ $\mathrm{P_A}$ and $\mathrm{K_A}$ are offset below and above the plane of the fabric, respectively.
    While the $a=+1$ and $a=-1$ arrangements are topologically equivalent,  once the fabric deforms to one state (eg. $a=+1$), the zigzag yarn are tightly interlocked, and shifting to the other sheared state ($a=-1$) becomes difficult. 
    In the sheared state, the fabric shrinks in the wale direction and thickens. 
    Despite the increase in thickness, the sheared \textbf{wvy} fabric may exhibit limited extensibility under tensile forces since the tightly packed yarn restrict motion and undergo significant strain.
    The sheared states belong to $p112/a\, (\text{L7})$, the maximal t-subgroup with monoclinic/oblique lattice~\cite{supplmater}.
    
    Our other racemic fabric  (rightmost image Fig.~\ref{fig:Chiralweft}A) also has one unit of each handedness in its unit cell, and it belongs to $p112/a$ (L7). 
    The obliquely arranged $\mathrm{K_A}$s are always higher than $\mathrm{P_A}$s, and yarn segments joining them form domain boundaries. 
    The yarn in the domain boundaries is tilted in the thickness direction of the fabric; therefore, this racemic fabric is shorter than the enantiomeric \textbf{wvx}$\pm$ fabrics in their relaxed states (Fig.~\ref{fig:Chiralweft}B). 
    The domain boundaries facilitate an enhanced extensibility response of the fabric. The racemic \textbf{wvy} fabric also has domain boundaries and is shorter than the enantiomeric \textbf{wvx}$\pm$ fabrics. However, its rest state is still longer than the other racemic fabric due to the suppression of out-of-plane zig-zag deformations of the yarn due to frustration caused by the point defects.    

    %\textcolor{magenta}{chirality of stitch. Knit stitch and purl stitch form knits that can be untangled (?).}

%%%%%%%%%%%%%%%% REFERENCES %%%%%%%%%%%%%%%

\clearpage % Clear all remaining figures and tables then start a new page

% The list of references goes after the main text and before the acknowledgements
% When preparing an initial submission, we recommend you use BibTeX, like this:
%
%\bibliography{knit_knot_net} % for a file named

%\bibliographystyle{sciencemag}

% After the paper has completed peer review and been revised ready for acceptance,
% you should comment out the lines above and copy-paste the contents of your .bbl
% file here instead. This will help ensure that our conversion software works correctly.
% Remember to re-run BibTeX first - check the timestamp!
%
% Example of the first three entries copy-pasted from science_template.bbl:
%
%\begin{thebibliography}{1}
%
%\bibitem{example}
%A.~N. {Author}, An example reference. \emph{Journal of Improbable Research}
%  \textbf{1}, 67 (2020).
%
%\bibitem{example2}
%F.~M. {Surname}, S.~{Author}, A second example. \emph{Interesting Research
%  Letters} \textbf{32}, 897 (2019).
%
%\bibitem{example_preprint}
%P.~{One}, P.~{Two}, P.~{Three}, {An unpublished preprint}. \emph{preprint}
%  (2021), arXiv:2101.12345.
%
%\end{thebibliography}

%%%%%%%%%%%%%%%% ACKNOWLEDGEMENTS %%%%%%%%%%%%%%%

\section*{Acknowledgments}
    %We acknowledge XXX. We thank XXX for measurements of XXX using XXX and XXX for XXX.

\paragraph*{Funding:}
    This study was supported by the World Premier International Research Center Initiative Program, International Institute for Sustainability with Knotted Chiral Meta Matter (WPI-SKCM$^{\text{2}}$), MEXT, Japan; 
    by JST SPRING (JPMJSP2128) and JSPS Grant-in-Aid for JSPS Fellows (Grant Number JP25KJ2180) to S.T.; 
    and by JSPS KAKENHI (Grant Number JP25K07005) to Y.Kotorii.
%\textcolor{cyan}{List the grants, fellowships etc. that funded the research; use initials to specify which author(s) were supported by each source. Include grant numbers when appropriate or required by the funding agency. For example: F.~A. was funded by the Generous Science Agency grant~2372.}
\paragraph*{Author contributions:}
    Conceptualization: S.T., E.A.M., Y.Kotorii, and K.I.
    Methodology: S.T., K.Y., and K.I.
    Investigation: S.T., Y.Kochi, K.Y., E.A.M., and K.I.
    Visualization: S.T. and K.I.
    Funding acquisition: S.T., Y.Kotorii, and K.I.
    Project administration: T.A. and K.I.
    Supervision: K.I.
    Writing---original draft: S.T. and K.I.
    Writing---review \& editing: S.T., Y.Kochi, K.Y., E.A.M., and Y.Kotorii
    .
\begin{comment}
TENTATIVE
\begin{table}[h]
    \centering
    \begin{tabular}{ll}
    \hline
        CRediT  & Author Abbreviations  \\
    \hline
        Conceptualization   &   S.T., E.A.M., Y.Kotorii, K.I.   \\
        Methodology &   S.T., K.Y., K.I.\\
        Investigation   &   S.T., (Y.Kochi), K.Y., E.A.M., K.I.\\
        Visualization   &   S.T., K.I.\\
        Funding acquisition &   S.T., K.I.\\
        Project administration  &   T.A., K.I.\\
        Supervision &   K.I.\\
        Writing---original draft    &   S.T., K.I.\\
        Writing---review \& editing &   S.T., Y.Kochi, K.Y., E.A.M., Y.Kotorii, (T.A.)\\
    \hline
    \end{tabular}
\end{table}
\end{comment}
%\textcolor{cyan}{List each author’s contributions to the paper. Use initials to abbreviate author names.}
\paragraph*{Competing interests:}
    The authors declare that a Japanese patent application related to this work has been filed by Hiroshima University and Waseda University, with inventors K.I., S.T., and T.A.; the application number is withheld pending publication.
    %There are no competing interests to declare.
%\textcolor{cyan}{Disclose any potential conflicts of interest for all authors, such as patent applications, additional affiliations, consultancies, financial relationships etc. See the journal editorial policies web page for types of competing interest that must be declared. If there are no competing interests, state:``There are no competing interests to declare.''}
\paragraph*{Data and materials availability:}
    All data are available in the main text, the supplementary materials, or Zenodo\cite{repository}.
\subsection*{Supplementary materials}
    Materials and Methods\\
    Supplementary Text\\
    Figs.~\ref{fig:sup_hopfcg} to \ref{fig:sup_layergroup}\\
    %Tables S1 to S4\\
    References \textit{(3-\arabic{enumiv})}\\ % automatically fills out the last reference number
    % (filling out the other numbers automatically is possible but fiddly and liable to break)
    %Movie S1\\
    %Data S1

%%%%%%%%%%%%%%%% END OF MAIN TEXT %%%%%%%%%%%%%%%
\newpage

%%%%%%%%%%%%%%%% START OF SUPPLEMENT %%%%%%%%%%%%%%%
%%%%%%%%%%%%%%%% START OF SUPPLEMENT %%%%%%%%%%%%%%%

% Figures, tables, equations and pages in the supplement are numbered S1, S2 etc.
\renewcommand{\thefigure}{S\arabic{figure}}
\renewcommand{\thetable}{S\arabic{table}}
\renewcommand{\theequation}{S\arabic{equation}}
\renewcommand{\thepage}{S\arabic{page}}
\setcounter{figure}{0}
\setcounter{table}{0}
\setcounter{equation}{0}
\setcounter{page}{1} % not 0 as \newpage already started a supplementary page
% References continue the numbering from the main text.

%%%%%%%%%%%%%%%% SUPPLEMENT TITLE PAGE %%%%%%%%%%%%%%%

\begin{center}
\section*{Supplementary Materials for\\ \scititle}
% Author list for the supplement
% Indicate the corresponding authors, but do NOT include institutions here
% It would be nice if the template auto-generated this, but doing so is complicated...
    \sciauthor
    % we're not in a \author{} environment this time, so use \\ for a new line
    \\
    \sciemail
\end{center}

% Fill out the numbers for each type of supplementary material,
% and delete any lines that aren't applicable.
% These are just example numbers that don't match the rest of this template.
\subsubsection*{This PDF file includes:}
Materials and Methods\\
Supplementary Text\\
Figures S1 to S4\\
Tables S1\\
%Captions for Movies S1 to S2\\
%Captions for Data S1 to S2

%\subsubsection*{Other Supplementary Materials for this manuscript:}
%Movies S1 to S2\\
%Data S1 to S2

\newpage

%%%%%%%%%%%%%%%% MATERIALS AND METHODS %%%%%%%%%%%%%%%

\subsection*{Materials and Methods}

\headingsm{Fabrication of tubular fabrics}
    We used Edo Braided Cord (Handicraft Strings No. 40, GTIN 4549131928648) from DAISO INDUSTRIES CO., LTD, which is $100\%$ nylon yarn with $3\,\si{m}$ in length and $4\,\si{mm}$ in width, hereafter referred to as the ``nylon yarn''.

    We fabricated four types of tubular fabrics by hand: a pair of enantiomeric fabrics and two types of racemic fabrics. A handmade tubular fabric was made of two nylon yarns and supported by a metal ring at the each wale ends. The metal rings were Card ring -35mm-1.38$^{\prime\prime}$- (D-137 Card Ring 6373) from DAISO INDUSTRIES CO., LTD, which is made of steel. Each fabric sample consisted of 6 rows and were made at equal tensions and stitch size. 
    
%=ParametricCurves=============================================================================
\vspace{1ex}
\headingsm{Parametric curves illustrating yarns of knitted fabrics}
    Typical knitted fabrics are stockinette, rib, garter and seed. These three-dimensional structures were visualized with parametric curves. While these curves do not reproduce structures with minimum elastic energy, the topology is correctly represented. Parametric curves are defined as $f: \mathbb{R}\times\mathbb{Z} \to \mathbb{R}^2\times I; (t, n) \mapsto (f_x, f_y, f_z)$. When $(m_1,m_2)$ are taken as periods, $f$ induces a mapping from $\mathbb{R}/m_1\mathbb{Z} \times \mathbb{Z}/m_2\mathbb{Z}$ to $\mathbb{R}/m_1\mathbb{Z} \times \mathbb{R}/m_2\mathbb{Z} \times I \cong T^2 \times I$
    where $m_2$ is the number of yarns per unit cell.
    For all typical fabrics, $f_x$ and $f_y$ are common:
    \begin{equation}
        \begin{aligned}
            f_x &=  t   +   0.3 \sin(4\pi t)   ,\\
            f_y &=  n   +   0.9 \, c,
        \end{aligned}
    \end{equation}
    where $c \coloneqq \cos(2\pi t)$.
    Each characteristic appears in $f_z$:
    \vspace{-1\baselineskip}
    \begin{subequations}
        \begin{align}
        \intertext{for stockinette,}
        \vspace{-1\baselineskip}
        f_z &=  -0.5 \, c^2
        ;\\
        \intertext{for garter,}
        f_z &=  -0.5 (-1)^n \, c
        ;\\
        \intertext{for rib,}
        f_z &=  -0.25 \cos{(\pi t)} 
            \left(
                2.5 - c - c^2
            \right)
        ;\\
        \intertext{for seed,}
        f_z &=  -0.5 (-1)^n \, \cos{(\pi t)}
            \left(
                0.4 - c^2
            \right)
        .
        \end{align}
    \end{subequations}
    These $f$ above describe monodomain, domain boundaries and point defects with vortex numbers $\pm 4$. The topology of point defects with vortex numbers $\pm 2$ is given by
    \begin{equation}
        \begin{aligned}
            f_x &=  t   +   0.3 \sin(4\pi t)   ,\\
            f_y &=  n   +   
            \left(
                0.9 - 0.2\,\theta_n^+
            \right)
            \, c  ,\\
            f_z &= 
            0.2\,\theta_n^+
            \left(
                \left(
                    0.2 - (1 + c)^2 + 0.7(1 + c^2)^3
                \right)
                (1 + c)\cos{(\pi t)} 
                - c(1 - c)(1 + 2c)
            \right)
            \\&\hspace{1em}
            + 0.4\,\theta_n^-
            \left(
                2
                \left(
                    1 + 3(1 - c^3)^3
                \right)
                \cos{(\pi t)} 
                + 3c - (1 + c)^3)
            \right)
            ,
        \end{aligned}
    \end{equation}
    where $\theta_\bullet^\pm:\mathbb{Z}\to\{0,1\}$
    \begin{equation}
        \begin{aligned}
            \theta_n^+ = 1,
            &\hspace{1em}
            \theta_n^- = 0
            \hspace{1em}
            (n \equiv 0 \mod 2)
            ,\\
            \theta_n^+ = 0,
            &\hspace{1em}
            \theta_n^- = 1
            \hspace{1em}
            (n \equiv 1 \mod 2)
            .
        \end{aligned}
    \end{equation}

     Enantiomeric \textbf{wvx}$\pm$ fabrics are consist of interlocked helices:
     \begin{equation}
        \begin{aligned}
            f_x &=  t - 0.05 a n,\\
            f_y &=  0.17 n + 0.2 \cos{(2 \pi t + \pi n)},\\
            f_z &=  0.2 h \sin{ (2 \pi t + \pi n)},
        \end{aligned}
    \end{equation}
    where $h = \pm 1$ determines sense of helices, $a = 0, \pm 1$ represents direction of shear; $a = 0$ is initial deferomation-free state.

     \textbf{wvy} fabric is modeled as
     \begin{equation}
        \begin{aligned}
            f_x &=  t +0.25 a n + 0.1 |a| \sin{(4 \pi t)},\\
            f_y &=  0.2 n + (0.23 - 0.03 |a|) \cos{(2 \pi t + \pi n )} + 0.07 a \sin{(4 \pi t )},\\
            f_z &=  (1 - |a|) (-0.2 \sin{(4 \pi t + \pi n )}) + 0.3 a \cos{(2 \pi t)},
        \end{aligned}
    \end{equation}
    $a = 0$ is the initial state, $a=\pm1$ are the sheared states, and $0<|a|<1$ are their intermediate states.
    This structure is centrosymmetric because it is invariant under inversion. This transformation is realized by combining $t \mapsto - ( t + 1/2), n \mapsto -n$ and the following translation by $-1/2$ in the $x$-direction.

    For chainmails, two integers $m, n$ are used to represent each rings. \textbf{cmj} chainmail:
    \begin{equation}
        \begin{aligned}
            f_x &=  m + n + 0.8 \cos{(2 \pi s)},\\
            f_y &=  m - n + 0.8 \sin{(2 \pi s)},\\
            f_z &=  0.2 a \sin{(8\pi s)},
        \end{aligned}
    \end{equation}
    where $a=\pm 1$ determines the chirality of the chainmail.

    The historical 4-in-1 chainmail
    \begin{equation}
        \begin{aligned}
            f_x &=  m + n + 0.9 \cos{(2 \pi s)},\\
            f_y &=  \cos{a} (m - n + 0.9  \sin{(2 \pi s)}),\\
            f_z &=  0.9 (-1)^{m-n} \sin{a} \sin{(2\pi s)},
        \end{aligned}
    \end{equation}
    where $a$ is pitch angle of rings.

\headingsm{Elastic energy of curved yarn}
    Mechanical behaviour of yarns are described as Kirchhoff elastic rods~\cite{Patil2020sci,Langer1996siamreview}. For a uniform rod, we introduce the \textit{material frame} $\{e_{\text{r}}, e_{\text{p}}, e_{\text{y}}\}$ defined at each points. When $e_{\text{r}}$ is set along the tangent of the centerline curve, $e_{\text{r}}, e_{\text{p}}, e_{\text{y}}$ represent axes of roll, pitch and yaw, respectively. The material frame is defined to be constant over $s$ for a straight and untwisted rod. When the centreline is a straight line, $e_{\text{r}}$ is defined to be constant over $s$. For the untwisted rod, $\tau_{\text{r}}$ (defined later) is defined to be $0$. The material frame obeys ``Frenet equations'':
    \begin{equation}
        \frac{d}{ds}
        \left(
            \begin{array}{c}
                e_{\text{r}}    \\
                e_{\text{p}}    \\
                e_{\text{y}}
            \end{array}
        \right)
        =
        \left(
            \begin{array}{ccc}
                0   &\kappa_{\text{y}}   &-\kappa_{\text{p}}    \\
                -\kappa_{\text{y}}  &0  &\tau_{\text{r}}    \\
                \kappa_{\text{p}}   &-\tau_{\text{r}}    &0
            \end{array}
        \right)
        \left(
            \begin{array}{c}
                e_{\text{r}}    \\
                e_{\text{p}}    \\
                e_{\text{y}}
            \end{array}
        \right)
        ,
    \end{equation}
    where $\tau_{\text{r}}, \kappa_{\text{p}}, \kappa_{\text{y}}$ represent rotation rate per $s$ around the roll, pitch and yaw axes. The deformed rod has extra elastic energy $E$, which is within locally small deformation represented as 
    \begin{equation}
        U = \int ds\, \left(\frac{K_{\text{r}}}{2} {\tau_{\text{r}}}^2+\frac{K_{\text{p}}}{2} {\kappa_{\text{p}}}^2 + \frac{K_{\text{y}}}{2} {\kappa_{\text{y}}}^2\right)
        ,
    \end{equation}
    here we call $K_{\text{r}}, K_{\text{p}}, K_{\text{y}}$ as elastic constants. For the uniform rod, the axes $e_{\text{p}}, e_{\text{y}}$ are selected such that the elastic energy does not contain $\kappa_{\text{p}}\kappa_{\text{y}}$ terms. As the elastic constants are two orders of magnitude smaller than the Young's modulus~\cite{Patil2020sci}, yarn stretching is negligible, making bending and twisting deformation the dominant modes.

    Frenet--Serret frame $\{e_{\text{t}}, e_{\text{n}}, e_{\text{b}}\}$ is a well-known frame to represent a curve embedded in the three-dimensional Euclidean space $\mathbb{E}^3$. We assume a curve $\gamma:\mathbb{R}^1\to\mathbb{E}^3;s\mapsto\gamma(s)$ with $s$ is the arclength that parametrises the curve. Frenet--Serret frame obeys Frenet equations (Frenet--Serret formulae):
    \begin{equation}
        \frac{d}{ds}
        \left(
            \begin{array}{c}
                \hat{T}    \\
                \hat{N}    \\
                \hat{B}
            \end{array}
        \right)
        =
        \left(
            \begin{array}{ccc}
                0   &\kappa   &0    \\
                -\kappa  &0  &\tau    \\
                0   &-\tau    &0
            \end{array}
        \right)
        \left(
            \begin{array}{c}
                \hat{T}    \\
                \hat{N}    \\
                \hat{B}
            \end{array}
        \right)
        ,
    \end{equation}
    where $\hat{T}, \hat{N}, \hat{B}$ are the unit tangent vector, the unit normal vector and the unit binormal vector, respectively, defined as
    \begin{equation}
        \begin{aligned}
            \hat{T} &\coloneqq  \frac{d\gamma}{ds}  ,\\
            \hat{N} &\coloneqq  \left|\frac{d^2\gamma}{ds^2}\right|^{-1} \frac{d^2\gamma}{ds^2} ,\\
            \hat{B} &\coloneqq  \hat{T} \times \hat{N}
            .
        \end{aligned}
    \end{equation}
    Curvature $\kappa$ and torsion $\tau$ are defined as
    \begin{equation}
        \begin{aligned}
            \kappa  &\coloneqq \left|\frac{d^2\gamma}{ds^2}\right|    ,\\
            \tau    &\coloneqq \left|\frac{d^2\gamma}{ds^2}\right|^{-2} \text{det}\left(\frac{d\gamma}{ds}, \frac{d^2\gamma}{ds^2}, \frac{d^3\gamma}{ds^3}\right)
            .
        \end{aligned}
    \end{equation}

    We will find relationship between material frame and Frenet--Serret frame. Assume
    \begin{equation}
        \begin{aligned}
            \hat{T} &=  e_{\text{r}}  ,\\
            \hat{N} &=  \cos{\psi}\, e_{\text{p}} + \sin{\psi}\, e_{\text{y}}   ,\\
            \hat{B} &=  -\sin{\psi}\, e_{\text{p}} + \cos{\psi}\, e_{\text{y}}
            .
        \end{aligned}
    \end{equation}
    Rotaion of axes around $\hat{T}$ is indicated by $\psi$. We obtain
    \begin{equation}
        \begin{aligned}
            \kappa_{\text{p}}   &=  -\kappa \sin{\psi}  ,\\
            \kappa_{\text{y}}   &=  +\kappa \cos{\psi}  ,\\
            \tau_{\text{r}}  &=  +\tau - \frac{d\psi}{ds}
            .
        \end{aligned}
    \end{equation} 
    While the centerline $\gamma$ has two degrees of freedom ($\kappa, \tau$), the material frame has three degrees of freedom ($\kappa_{\text{p}}, \kappa_{\text{y}}, \tau_{\text{r}}$) because a yarn with a finite cross section may twist around the roll axis. The case $\tau_{\text{r}}\equiv0$ is known as the \textit{Bishop frame}, which is used for regulating snake-like robots without roll joints. The Bishop frame is capable of representing any arbitrary Frenet--Serret frame, which implies that the yarn does not necessarily have to be twisted to describe any (centerline) curve. Consequently, the elastic energy of the yarn is minimized when $\tau_{\text{r}}$ is nonzero, which is achieved by selecting an appropriate value for $\psi$. If the terminals of the yarn are not fixed, meaning they can rotate freely without any restrictions, the contributions from bending rigidity must be taken into account. In the simplified case where $B\coloneqq K_{\text{p}} = K_{\text{y}}$ is assumed, the elastic energy becomes
    \begin{equation}
        U = \int ds\, \frac{B}{2} {\kappa}^2
        .
    \end{equation}
    
    Let $f$ be a curve that is parametrised by an arbitrary variable $t$, and consider a relationship $\Vec{\gamma}(s) = f(t)$ with a arclength parameter $s$:    
    \begin{equation}
        \begin{aligned}
            s   &=  \int^t dt\, \left|\frac{d f}{d t}\right|   ,\\
            \frac{d \gamma}{d s}
            &=
            \left| \frac{d f}{d t} \right|^{-1} \frac{d f}{d t}
            ,\\
            \frac{d^2 \gamma}{d s^2}
            &=
            \left| \frac{d f}{d t} \right|^{-2}
            \left(
                \frac{d^2 f}{d t^2}
                - \left| \frac{d f}{d t} \right|^{-2}
                \left( \frac{d f}{d t}\cdot\frac{d^2 f}{d t^2} \right)
                \frac{d f}{d t}
            \right)
            ,\\
            U &= \int ds\, \frac{B}{2} \left| \frac{d^2}{ds^2} \Vec{\gamma} (s) \right|^2
            = \int dt\, \frac{B}{2} \left| \frac{df}{dt} \right|^{-5} \left| \frac{d^2f}{dt^2} \times \frac{df}{dt} \right|^2
            ,
        \end{aligned}
    \end{equation}
    the expression of the elastic energy is identical to that in~\cite{Singal2024natcommun} and non-stretchable twist-free condition in~\cite{Patil2020sci}.
    Force acting on the yarn $F$ is given by
    \begin{equation}
        \begin{aligned}
            F
            & \coloneqq
            - \frac{\delta U}{\delta \gamma}
            \\
            &=
            - \sum_{n \ge 0} (-1)^n \frac{d^n}{ds^n} \frac{\partial u}{\partial \gamma^{(n)}} = - B \frac{d^4 \gamma}{ds^4}
            ,
        \end{aligned}
    \end{equation}
    where $u$ is the density of elastic energy of yarn per arclength.

%%%%%%%%%%%%%%%% SUPPLEMENTARY TEXT %%%%%%%%%%%%%%%

\subsection*{Supplementary Text}
%The Supplementary Text section can only be used to directly support statements made in the main text e.g. to present more detailed justifications of assumptions, investigate alternative scenarios, provide extended acknowledgements etc. Material in this section cannot claim results or conclusions that weren't mentioned in the main text. To refer to this section from the main text, just write (Supplementary Text).
%\subsubsection*{Example supplement heading}
%The two main sections of the supplement can be split up using headings.

% If your supplement is very short you might need to uncomment the following line to avoid
% layout problems with the figures and tables.
%\newpage

\begin{figure}
	\centering
	\includegraphics[width=0.67\textwidth]{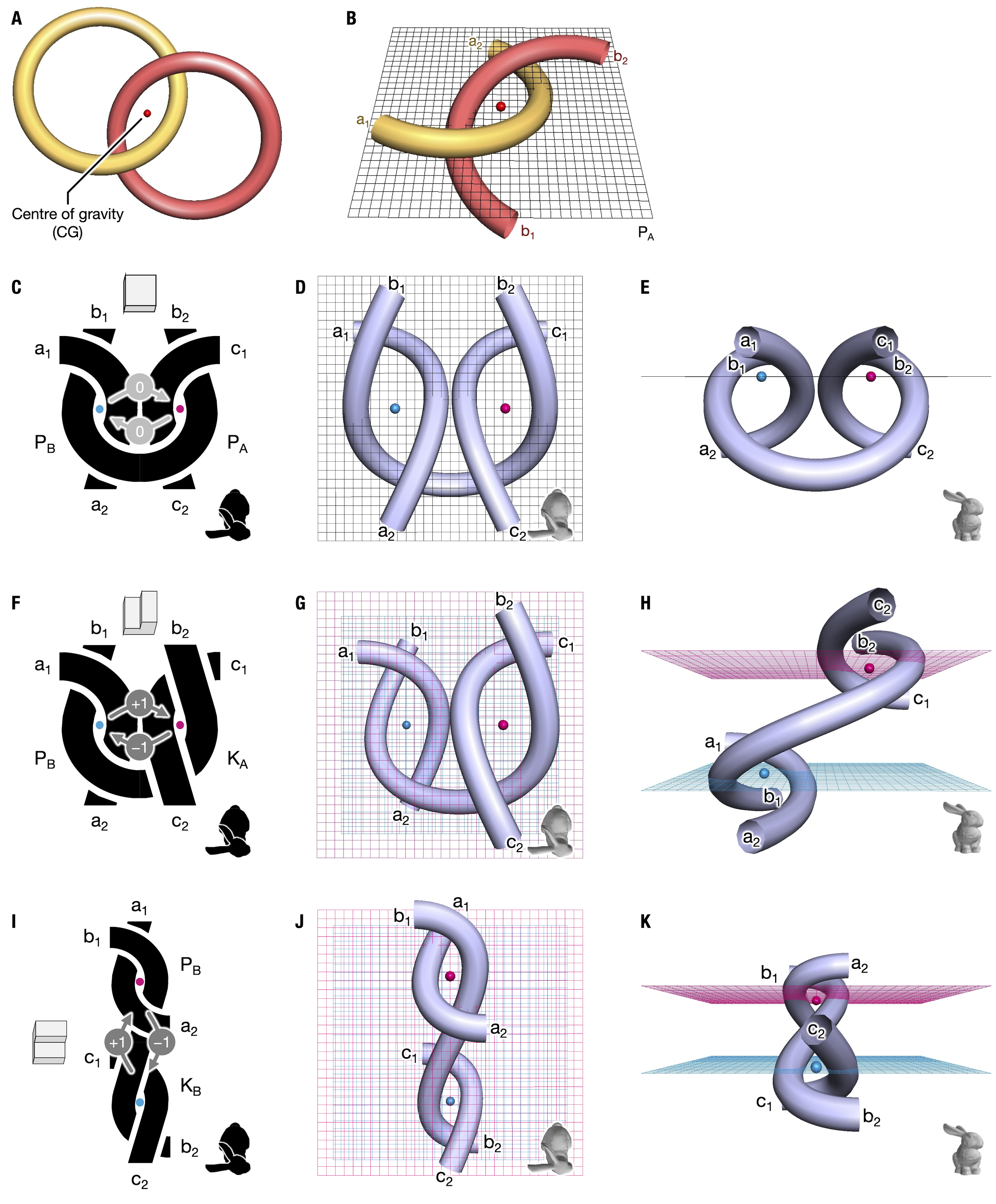}	
    
	\caption{\textbf{Center of gravity of a unit.}
        (\textbf{A}) Hopf link and (\textbf{B}) unit $\mathrm{P_A}$. The red point and the mesh indicate CG and the fabric plane, respectively.
        (\textbf{C}--\textbf{K}) diagram and three-dimensional configuration of unit pairs. The Stanford Bunny represents the angle of view. The red and blue points (CG of unit) are on the mesh with the same color.
    }
	\label{fig:sup_hopfcg}
\end{figure}

%=Block==================================================================================
\headingsm{Building Block}
    The concept of the center of gravity (CG) of a unit and the step between adjacent units was introduced. Consider a Hopf link formed by interlocking planar circles such that the centers of each circle lie at the intersections of their respective circular planes (Fig.~\ref{fig:sup_hopfcg}A). This Hopf link has its CG at the red point. Figure~\ref{fig:sup_hopfcg}B shows a perspective view of the $\mathrm{P_A}$ unit, which is the segments of the Hopf link that surround CG. The red point indicates CG of the unit, which is on the fabric plane represented by the black mesh. The unit has four arms: $\mathrm{a_1}$ and $\mathrm{b_2}$ are above the fabric plane, while $\mathrm{a_2}$ and $\mathrm{b_1}$ are below it. This configuration defines the relative height of each arm and CG.

    A step is defined between two adjacent units. Figure~\ref{fig:sup_hopfcg}C--E shows $\mathrm{P_B}\mathrm{P_A}$ neighbors in the course direction. CGs of the units are represented by blue and red points respectively, and are in the same plane. Figure~\ref{fig:sup_hopfcg}F--H shows $\mathrm{P_B}\mathrm{K_A}$ neighbors in the course direction. CGs of each unit are on different levels, with $\mathrm{K_A}$ being one step higher than $\mathrm{P_B}$. The units are connected by a segment $\mathrm{b_1}$--$\mathrm{b_2}$. Figures~\ref{fig:sup_hopfcg}I--K show $\mathrm{P_B}\mathrm{K_B}$ neighbors in the wale direction. Their CGs are on different levels, and $\mathrm{P_B}$ is one step higher than $\mathrm{K_B}$. 

%=LayerGroup=============================================================================
\vspace{1ex}
\headingsm{Assignment to Layer Group}
    The layer group $G_{3,2}$ delineates the symmetry of a two-dimensional plane, distinguishing between the front and back. It is a subgroup of the crystallographic space group $G_{3,3}$ and a supergroup of the plane group $G_{2,2}$, with an order of 80, which is intermediate between the space group $G_{3,3}$ (320 orders) and the wallpaper group $G_{2,2}$ (17 orders). The layer group is generated by augmenting the symmetry elements of the plane groups. Additional elements of order 2 within the plane, such as inversion centers, mirrors, glide planes, and two-fold axes, are introduced to all elements of the plane groups. Every layer group type has an equivalent space group type, which is constructed by incorporating translational symmetry normal to the fabric plane. The space group type equivalent to $p211\, (\text{L8})$ is $P2\, (\text{\#3})$. Some layer group types correspond to multiple plane group types, due to the various possible choices in symmetry reduction. The first approach involves projecting the three-dimensional symmetry operations of the layer group onto the two-dimensional symmetry operations of the corresponding plane group~\cite{IT_E2006}. This approach yields `Plane group 1', as shown in Tab.~\ref{tab:sup_example}. Under this projection, an inversion center at a point P is mapped to a two-fold rotation axis normal to the layer plane at point P. Additionally, a mirror plane perpendicular to the $a$-axis is mapped to a mirror line that is perpendicular to the same axis. The second approach extracts the intrinsically two-dimensional symmetry operations from the layer group. The resulting symmetry is shown as `Plane Group 2' (Tab.~\ref{tab:sup_example}); in this case, $c222\, (\text{L22})$ is reduced to $p2\, (\text{P2})$. In this reduction, the two-fold rotation axis perpendicular to the fabric plane is retained, as it is compatible with plane symmetry. In contrast, the two-fold rotation about the $a$-axis is discarded, as it represents a symmetry operation extending beyond the plane. Of course, Plane Group 2 is a subgroup of the  corresponding layer group.

    Among 80 types of layer group, 17 of them form chiral crystals. Rigorously, the 17 types are \textit{Sohncke} types, which are not exactly chiral as space groups but are only invariant under operations of the first kind that preserve handedness. The operations of the second kind, i.e., reflection, glide, and inversion, invert the handedness. Three-dimensional space groups (respectively, wallpaper groups) have 65 types (5 types) of Sohncke groups~\cite{Nespolo2021japplcrystallogr}. Sohncke groups are divided into 11 enantiomorphic pairs of \textit{chiral space groups} and other achiral groups with 43 orders in the three-dimensional space. The chiral space groups themselves are chiral, meaning that their mirror images are distinct space groups. Chiral space group type $P3_1 21 \, (\text{IUCr number: }\# 152, \text{Fibrifold: }(3_1 \mathord{\ast} 3_1), \text{Point group: }D_3)$ occuring in $\ensuremath{\alpha}$-quartz has the enantiomorphic counterpart $P3_2 21 \, (\# 154, (3_1 \mathord{\ast} 3_1), D_3)$, which is a different type. A chiral space group always contains a screw axis $n_p$ or $n_{n-p}$ with $p \neq n/2$, where $n,p \in \mathbb{N}$, and these screw axes are distinguished as left- or right-handed rotations. Screw $n_{n/2}$ yields the identical result irrespective of the direction of rotation. Wallpaper group cannot contain screw axes; consequently, all 5 Sohncke types are achiral as space group. Since a crystal belonging to a Sohncke type is non-superimposable to its mirror image, such a crystal always has a chiral structure. An achiral Sohncke type $P 2_1 \, (\# 4, (2_1 2_1 2_1 2_1), C_2)$ occurs in thalidemide (TD) with enantiopure molecules, and crystals of (\textit{R})-TD and (\textit{S})-TD are distinguished~\cite{Matsumoto2025jacs}. A famous magnet $\mathrm{Cr Nb_3 S_6}$ belongs to an achiral Sohncke type of space group $P 6_3 22 \, (\# 182, (\mathord{\ast} 6_3 3_0 2_1), D_6)$~\cite{Togawa2012prl} but helical arrangement of spins is formed depending on the handedness of the crystal. Layer group may contain screw axes $n_{n/2}$, and have achiral 17 Sohncke types. 
    
\begin{figure}
	\centering
	\includegraphics[width=1.0\textwidth]{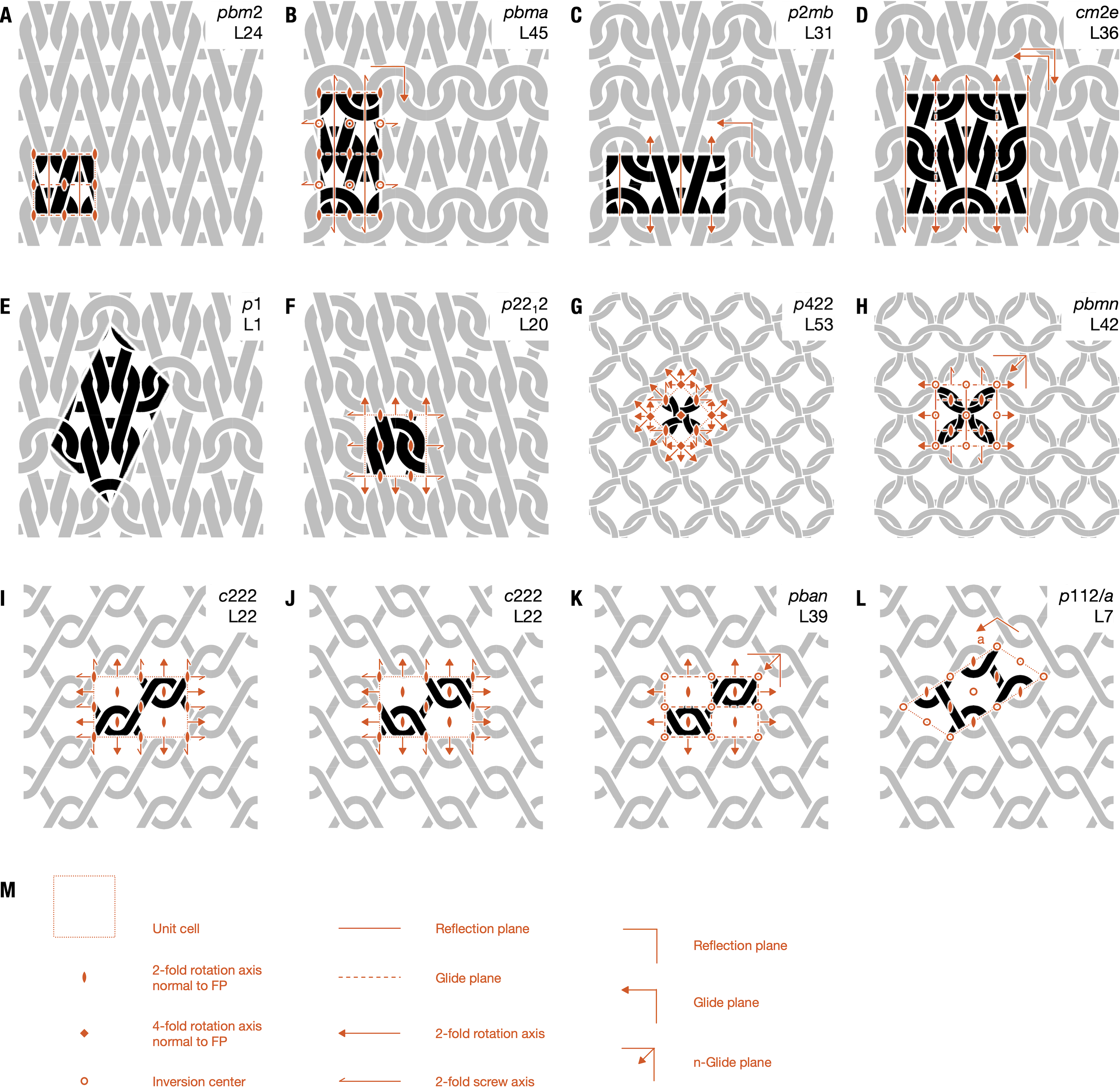}	

	\caption{\textbf{Layer group and symmetry elements. }
        (\textbf{A})~Stockinette, 
        (\textbf{B})~garter, 
        (\textbf{C})~$1\times 1$~rib,
        (\textbf{D})~seed. 
        (\textbf{G})~\textbf{cmj} chainmail,
        (\textbf{H})~4-in-1 chainmail, 
        (\textbf{I})~\textbf{wvx}$-$ fabric, 
        (\textbf{J})~\textbf{wvx}$+$ fabric,
        (\textbf{K})~\textbf{wvy} fabric,
        (\textbf{L})~racemic fabric\,2. 
        (\textbf{M})~notation of symmetry elements.
    }
	\label{fig:sup_layergroup}
\end{figure}

    Unlike crystallographic structure analysis techniques, such as X-ray diffraction, for which modules for space-group-type assignment are available, no such modules are, to our knowledge, accessible for the assignment of space groups to structures of fabrics. Therefore, we manually assigned the space group. First, we identified the unit cell of the fabric, which serves as the repeating unit. Next, we systematically examined the possible symmetry elements in the fabric. We compared the obtained unit cell and symmetry elements with layer group listed in \textit{International Tables for Crystallography} (\textit{IT})~\cite{IT_E2006}, and identified the layer group type of the fabric. The symmetry elements of the fabric shown in the paper are all listed in Fig.~\ref{fig:sup_layergroup}.

\begin{figure}
	\centering
	\includegraphics[width=1.0\textwidth]{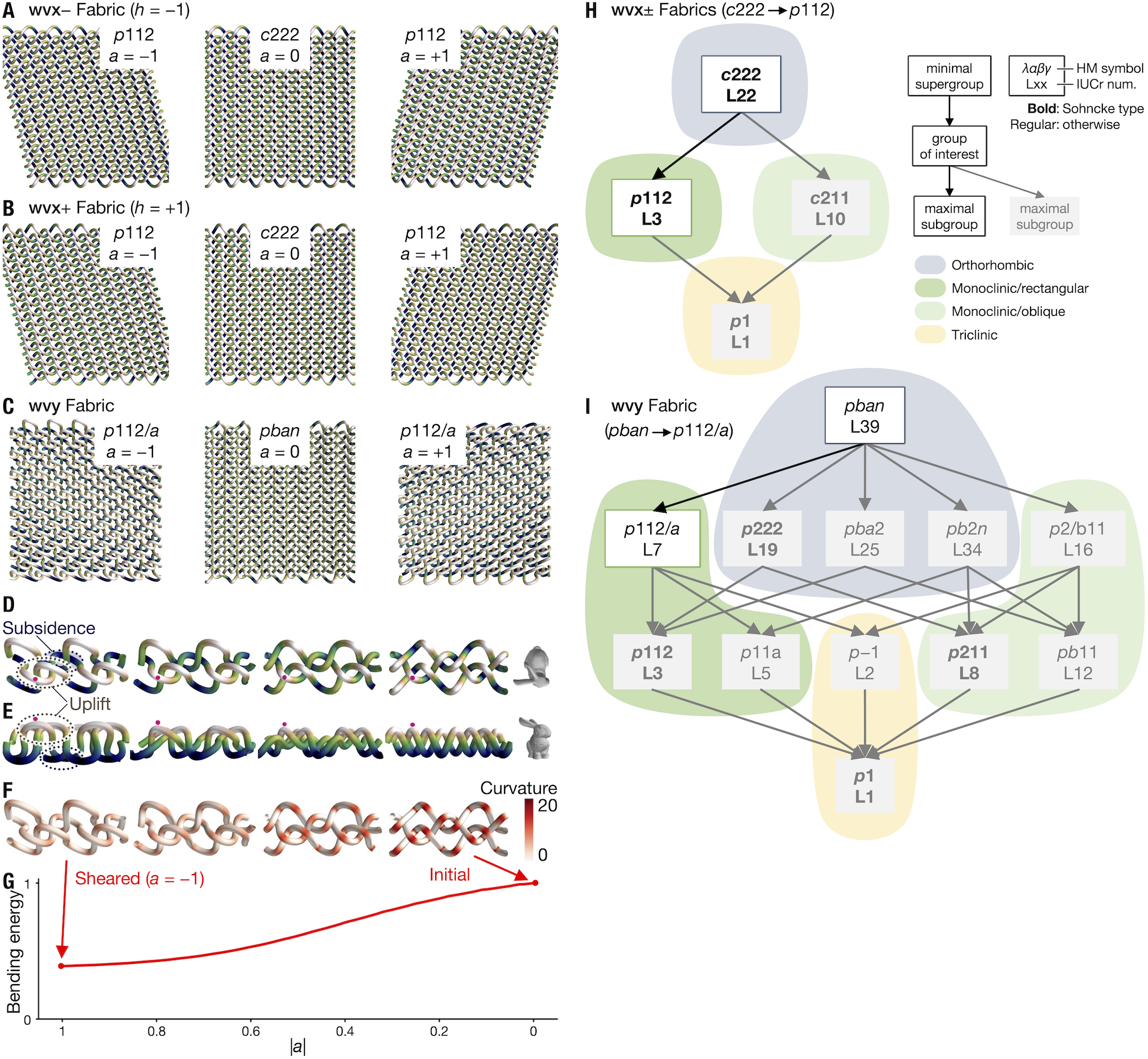}	
    
	\caption{\textbf{Shear deformation in knit-inspired fabrics.}
        (\textbf{A} and \textbf{B}) \textbf{wvx}$\pm$ fabrics. (\textbf{C}) \textbf{wvy} fabric. For \textbf{wvy} fabric, (\textbf{D} and \textbf{E}) show continuous deformation between shear-deformed and initial sates, (\textbf{F}) shows distribution of curvature of the yarns. (\textbf{G}) displays bending energy per unit cell, which is normalized by yarn length and the unity corresponds to non-deformed state. The Stanford Bunny represents the angle of view. (\textbf{H} and \textbf{I}) Group--subgroup relation in layer group types of \textbf{wvx}$\pm$ and \textbf{wvy} fabrics, respectively. Subgroup and supergroup refer to \textit{translationengleiche} (t-) relations. The directed graphs were created with reference to \textit{IT}~\cite{IT_E2006}. HM symbol referrers to Hermann--Mauguin symbol for layer group types. The supergroup types of $c222$ and $pban$ are not shown here, although they exist.
    }
	\label{fig:sup_weft_deformation}
\end{figure}

    Figure~\ref{fig:sup_weft_deformation} shows three-dimensional structure of fabrics in Fig.~\ref{fig:Chiralweft}. The initial and sheared states are topologically equivalent because these two states have infinite intermediate states, which is represented by the deformation parameter $a$. The enantiomeric \textbf{wvx}$\pm$ fabrics crystallize in $c222$ and $p112$ at the initial and sheared states, respectively (Fig.~\ref{fig:sup_weft_deformation}A and B). $c222$ has 2 maximal subgroup types $p122$ and $c211\, (\text{L10})$. Among them, only $p112$ has oblique lattice system, which is selected in the sheared state (Fig.~\ref{fig:sup_weft_deformation}H). For \textbf{wvy} fabric, $pban$ has 5 maximal subgroup types, including $p112/a$, which is the only one exhibiting an oblique lattice system (Fig.~\ref{fig:sup_weft_deformation}I).
    Gradual winding in the deformed state reduces curvature and then elastic energy, which leads to the spontaneous deformation. Some literatures discussed layer group of chainmails~\cite{Liu2018chemsocrev,Frank2020bridges}

\begin{figure}
	\centering
	\includegraphics[width=1.0\textwidth]{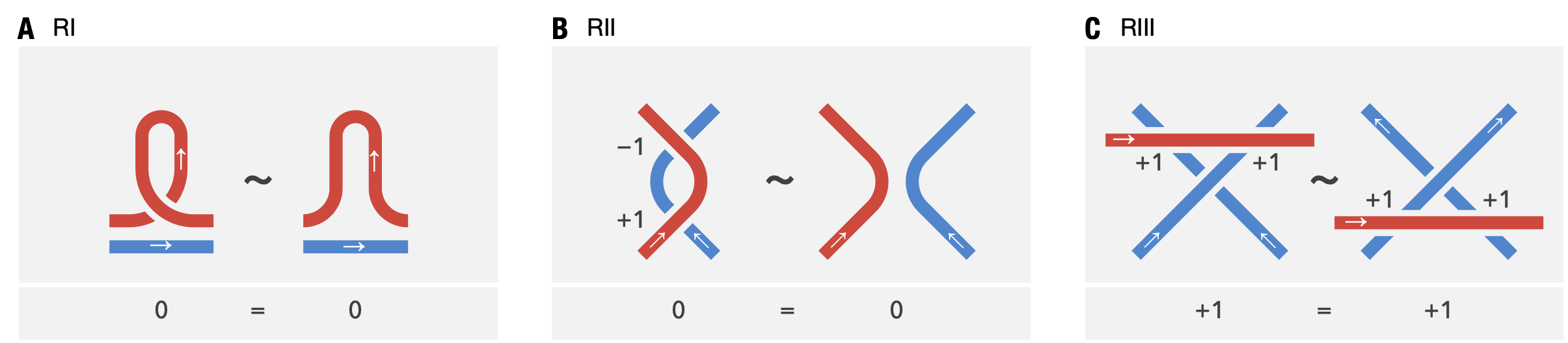}	
    
	\caption{\textbf{Reidemeister moves.} 
    }
	\label{fig:sup_linkingnumber}
\end{figure}

%=LinkNum=================================================================================
\vspace{1ex}\headingsm{Linking Number for Doubly Periodic Tangles}
    We defined the linking number per unit cell $Lk$. While the linking number is usually defined in the three-dimensional sphere $S^3$, $Lk$ here is defined in a thickened torus $T^2 \times I$, therefore, the term `linking number $Lk$' in this paper refers to the one per unit cell. The direction of the yarns was determined from left to right. Weft yarn meander in the direction of the wale, yet proceed in the course direction as a whole. In the fabrics under consideration in this paper, only 2-component links or a knot were observed. These components are designated as 1 and 2. $Lk$ is computed from the sign of crossing $c$ of $D_1$ and $D_2$. $\mathrm{sign}(c)$ is a binary defined to be $\mathrm{sign}(c) = +1$ if the overpass goes through from top to bottom of the underpass; otherwise, $\mathrm{sign}(c) = -1$ (Fig.~\ref{fig:Chiralweft}).  
    The linking number $Lk$ remains invariant under Reidemeister moves in $T^2 \times I$ (Fig.~\ref{fig:sup_linkingnumber}). Since Reidemeister moves are local continuous deformation, these transformations yield equivalent results in $S^3$.

%=Vortex=================================================================================
%\vspace{1ex}
%\headingsm{Vortex as Chiral Topological Invariant}

%%%%%%%%%%%%%%%% SUPPLEMENTARY FIGURES %%%%%%%%%%%%%%%

%%%%%%%%%%%%%%%% SUPPLEMENTARY TABLES %%%%%%%%%%%%%%%	
\setlength{\tabcolsep}{2pt}
\small
\begin{longtable}{llcrlllrllrlllcrlllc}
% Captions go above tables
	\caption{\textbf{All captions must start with a short bold sentence, acting as a title.}
        The sequential numbering of layer group, space group and plane group types listed in Refs.~\cite{IntTabCryst2016A} and \cite{IT_E2006} is represented by L, \# and P, respectively. The point group represents the geometric crystal class of each layer group. Horizontal lines separate layer groups according to their crystal class/lattice system, as indicated by their corresponding point groups. HM, Sf and Of stand for Hermann--Mauguin, Sch\"{o}nflies and orbifold~\cite{Conway2008symmetry} notation, respectively. Sk indicates whether the group is of Sohncke type, with $+$ for a Sohncke type and $-$ otherwise. 
        %Follow the same style as main text tables. If the design is similar to previous tables, avoid repetition by refering back to them.
        }
	\label{tab:sup_example} % give each table a logical label name
    \\
    \hline \hline
    \multicolumn{3}{l}{Layer group} &&\multicolumn{3}{l}{Point group} &&\multicolumn{2}{l}{Space group} &&\multicolumn{4}{l}{Plane group 1} &&\multicolumn{4}{l}{Plane group 2} \\
    \cline{1-3}\cline{5-7}\cline{9-10}\cline{12-15}\cline{17-20}
    L & HM & Sk && HM & Sf & Of &&\# & HM &&P & HM & Of & Sk &&P & HM & Of & Sk \\\hline
    \endfirsthead
    
    \hline \hline
    \multicolumn{3}{l}{Layer group} &&\multicolumn{3}{l}{Point group} &&\multicolumn{2}{l}{Space group} &&\multicolumn{4}{l}{Plane group 1} &&\multicolumn{4}{l}{Plane group 2} \\
    \cline{1-3}\cline{5-7}\cline{9-10}\cline{12-15}\cline{17-20}
    L & HM & Sk && HM & Sf & Of &&\# & HM &&P & HM & Of & Sk &&P & HM & Of & Sk \\\hline
    \endhead

\hline \hline
\endfoot

1& $p1$ & $+$ && $1$ & $C_1$ & $11$ && 1 & $P1$ && 1 & $p1$ & $\circ$ & $+$ && 1 & $p1$ & $\circ$ & $+$ \\
2& $p\bar{1}$ & $-$ && $\bar{1}$ & $C_i$ & $\mathord{\times}$ && 2 & $P\bar{1}$ && 2 & $p2$ & $2222$ & $+$ && 1 & $(p1)$ & $\circ$ & $+$ \\
\hline 3& $p112$ & $+$ && $2$ & $C_2$ & $22$ && 3 & $P2$ && 2 & $p2$ & $2222$ & $+$ && 2 & $p2$ & $2222$ & $+$ \\
4& $p11m$ & $-$ && $m$ & $C_s$ & $\mathord{\ast}$ && 6 & $Pm$ && 1 & $p1$ & $\circ$ & $+$ && 1 & $(p1)$ & $\circ$ & $+$ \\
5& $p11a$ & $-$ && $m$ & $C_s$ & $\mathord{\ast}$ && 7 & $Pa$ && 1 & $p1$ & $\circ$ & $+$ && 1 & $(p1)$ & $\circ$ & $+$ \\
6& $p112/m$ & $-$ && $2/m$ & $C_{2h}$ & $2\mathord{\ast}$ && 10 & $P2/m$ && 2 & $p2$ & $2222$ & $+$ && 2 & $(p2)$ & $2222$ & $+$ \\
7& $p112/a$ & $-$ && $2/m$ & $C_{2h}$ & $2\mathord{\ast}$ && 13 & $P2/a$ && 2 & $p2$ & $2222$ & $+$ && 2 & $(p2)$ & $2222$ & $+$ \\
\hline 8& $p211$ & $+$ && $2$ & $C_2$ & $22$ && 3 & $P2$ && 3 & $pm$ & $\mathord{\ast} \mathord{\ast}$ & $-$ && 1 & $(p1)$ & $\circ$ & $+$ \\
9& $p2_1 11$ & $+$ && $2$ & $C_2$ & $22$ && 4 & $P2_1$ && 4 & $pg$ & $\mathord{\times} \mathord{\times}$ & $-$ && 1 & $(p1)$ & $\circ$ & $+$ \\
10& $c211$ & $+$ && $2$ & $C_2$ & $22$ && 5 & $C2$ && 5 & $cm$ & $\mathord{\ast} \mathord{\times}$ & $-$ && 1 & $(p1)$ & $\circ$ & $+$ \\
11& $pm11$ & $-$ && $m$ & $C_s$ & $\mathord{\ast}$ && 6 & $Pm$ && 3 & $pm$ & $\mathord{\ast} \mathord{\ast}$ & $-$ && 3 & $pm$ & $\mathord{\ast} \mathord{\ast}$ & $-$ \\
12& $pb11$ & $-$ && $m$ & $C_s$ & $\mathord{\ast}$ && 7 & $Pa$ && 4 & $pg$ & $\mathord{\times} \mathord{\times}$ & $-$ && 4 & $pg$ & $\mathord{\times} \mathord{\times}$ & $-$ \\
13& $cm11$ & $-$ && $m$ & $C_s$ & $\mathord{\ast}$ && 8 & $Cm$ && 5 & $cm$ & $\mathord{\ast} \mathord{\times}$ & $-$ && 5 & $cm$ & $\mathord{\ast} \mathord{\times}$ & $-$ \\
14& $p2/m11$ & $-$ && $2/m$ & $C_{2h}$ & $2\mathord{\ast}$ && 10 & $P2/m$ && 6 & $p2mm$ & $\mathord{\ast} 2222$ & $-$ && 3 & $(pm)$ & $\mathord{\ast} \mathord{\ast}$ & $-$ \\
15& $p2_1/m11$ & $-$ && $2/m$ & $C_{2h}$ & $2\mathord{\ast}$ && 11 & $P2_1 /m$ && 7 & $p2mg$ & $22 \mathord{\ast}$ & $-$ && 3 & $(pm)$ & $\mathord{\ast} \mathord{\ast}$ & $-$ \\
16& $p2/b11$ & $-$ && $2/m$ & $C_{2h}$ & $2\mathord{\ast}$ && 13 & $P2/b$ && 7 & $p2mg$ & $22 \mathord{\ast}$ & $-$ && 4 & $(pg)$ & $\mathord{\times} \mathord{\times}$ & $-$ \\
17& $p2_1/b11$ & $-$ && $2/m$ & $C_{2h}$ & $2\mathord{\ast}$ && 14 & $P2_1 /b$ && 8 & $p2gg$ & $22 \mathord{\times}$ & $-$ && 4 & $(pg)$ & $\mathord{\times} \mathord{\times}$ & $-$ \\
18& $c2/m11$ & $-$ && $2/m$ & $C_{2h}$ & $2\mathord{\ast}$ && 12 & $C2/m$ && 9 & $c2mm$ & $2 \mathord{\ast} 22$ & $-$ && 5 & $(cm)$ & $\mathord{\ast} \mathord{\times}$ & $-$ \\
\hline 19& $p222$ & $+$ && $222$ & $D_2$ & $222$ && 16 & $P222$ && 6 & $p2mm$ & $\mathord{\ast} 2222$ & $-$ && 2 & $(p2)$ & $2222$ & $+$ \\
20& $p2_1 22$ & $+$ && $222$ & $D_2$ & $222$ && 17 & $P2_1 22$ && 7 & $p2mg$ & $22 \mathord{\ast}$ & $-$ && 2 & $(p2)$ & $2222$ & $+$ \\
21& $p2_1 2_1 2$ & $+$ && $222$ & $D_2$ & $222$ && 18 & $P2_1 2_1 2$ && 8 & $p2gg$ & $22 \mathord{\times}$ & $-$ && 2 & $(p2)$ & $2222$ & $+$ \\
22& $c222$ & $+$ && $222$ & $D_2$ & $222$ && 21 & $C222$ && 9 & $c2mm$ & $2 \mathord{\ast} 22$ & $-$ && 2 & $(p2)$ & $2222$ & $+$ \\
23& $pmm2$ & $-$ && $mm2$ & $C_{2v}$ & $\mathord{\ast} 22$ && 25 & $Pmm2$ && 6 & $p2mm$ & $\mathord{\ast} 2222$ & $-$ && 6 & $p2mm$ & $\mathord{\ast} 2222$ & $-$ \\
24& $pma2$ & $-$ && $mm2$ & $C_{2v}$ & $\mathord{\ast} 22$ && 28 & $Pma2$ && 7 & $p2mg$ & $22 \mathord{\ast}$ & $-$ && 7 & $p2mg$ & $22 \mathord{\ast}$ & $-$ \\
25& $pba2$ & $-$ && $mm2$ & $C_{2v}$ & $\mathord{\ast} 22$ && 32 & $Pba2$ && 8 & $p2gg$ & $22 \mathord{\times}$ & $-$ && 8 & $p2gg$ & $22 \mathord{\times}$ & $-$ \\
26& $cmm2$ & $-$ && $mm2$ & $C_{2v}$ & $\mathord{\ast} 22$ && 35 & $Cmm2$ && 9 & $c2mm$ & $2 \mathord{\ast} 22$ & $-$ && 9 & $c2mm$ & $2 \mathord{\ast} 22$ & $-$ \\
27& $pm2m$ & $-$ && $(m2m)$ & $C_{2v}$ & $\mathord{\ast} 22$ && 25 & $Pm2m$ && 3 & $pm$ & $\mathord{\ast} \mathord{\ast}$ & $-$ && 3 & $(pm)$ & $\mathord{\ast} \mathord{\ast}$ & $-$ \\
28& $pm2_1 b$ & $-$ && $(m2m)$ & $C_{2v}$ & $\mathord{\ast} 22$ && 26 & $Pm2_1 b$ && 3 & $pm$ & $\mathord{\ast} \mathord{\ast}$ & $-$ && 3 & $(pm)$ & $\mathord{\ast} \mathord{\ast}$ & $-$ \\
29& $pb2_1 m$ & $-$ && $(m2m)$ & $C_{2v}$ & $\mathord{\ast} 22$ && 26 & $Pb 2_1 m$ && 4 & $pg$ & $\mathord{\times} \mathord{\times}$ & $-$ && 4 & $(pg)$ & $\mathord{\times} \mathord{\times}$ & $-$ \\
30& $pb2b$ & $-$ && $(m2m)$ & $C_{2v}$ & $\mathord{\ast} 22$ && 27 & $Pb2b$ && 3 & $pm$ & $\mathord{\ast} \mathord{\ast}$ & $-$ && 4 & $(pg)$ & $\mathord{\times} \mathord{\times}$ & $-$ \\
31& $pm2a$ & $-$ && $(m2m)$ & $C_{2v}$ & $\mathord{\ast} 22$ && 28 & $Pm2a$ && 3 & $pm$ & $\mathord{\ast} \mathord{\ast}$ & $-$ && 3 & $(pm)$ & $\mathord{\ast} \mathord{\ast}$ & $-$ \\
32& $pm2_1 n$ & $-$ && $(m2m)$ & $C_{2v}$ & $\mathord{\ast} 22$ && 31 & $Pm2_1 n$ && 4 & $pg$ & $\mathord{\times} \mathord{\times}$ & $-$ && 3 & $(pm)$ & $\mathord{\ast} \mathord{\ast}$ & $-$ \\
33& $pb2_1 a$ & $-$ && $(m2m)$ & $C_{2v}$ & $\mathord{\ast} 22$ && 29 & $Pb 2_1 a$ && 4 & $pg$ & $\mathord{\times} \mathord{\times}$ & $-$ && 4 & $(pg)$ & $\mathord{\times} \mathord{\times}$ & $-$ \\
34& $pb2n$ & $-$ && $(m2m)$ & $C_{2v}$ & $\mathord{\ast} 22$ && 30 & $Pb2n$ && 5 & $cm$ & $\mathord{\ast} \mathord{\times}$ & $-$ && 4 & $(pg)$ & $\mathord{\times} \mathord{\times}$ & $-$ \\
35& $cm2m$ & $-$ && $(m2m)$ & $C_{2v}$ & $\mathord{\ast} 22$ && 38 & $Cm2m$ && 5 & $cm$ & $\mathord{\ast} \mathord{\times}$ & $-$ && 5 & $(cm)$ & $\mathord{\ast} \mathord{\times}$ & $-$ \\
36& $cm2e$ & $-$ && $(m2m)$ & $C_{2v}$ & $\mathord{\ast} 22$ && 39 & $Cm2e$ && 3 & $pm$ & $\mathord{\ast} \mathord{\ast}$ & $-$ && 5 & $(cm)$ & $\mathord{\ast} \mathord{\times}$ & $-$ \\
37& $pmmm$ & $-$ && $mmm$ & $D_{2h}$ & $\mathord{\ast} 222$ && 47 & $Pmmm$ && 6 & $p2mm$ & $\mathord{\ast} 2222$ & $-$ && 6 & $(p2mm)$ & $\mathord{\ast} 2222$ & $-$ \\
38& $pmaa$ & $-$ && $mmm$ & $D_{2h}$ & $\mathord{\ast} 222$ && 49 & $Pmaa$ && 6 & $p2mm$ & $\mathord{\ast} 2222$ & $-$ && 7 & $(p2mg)$ & $22 \mathord{\ast}$ & $-$ \\
39& $pban$ & $-$ && $mmm$ & $D_{2h}$ & $\mathord{\ast} 222$ && 50 & $Pban$ && 9 & $c2mm$ & $2 \mathord{\ast} 22$ & $-$ && 8 & $(p2gg)$ & $22 \mathord{\times}$ & $-$ \\
40& $pmam$ & $-$ && $mmm$ & $D_{2h}$ & $\mathord{\ast} 222$ && 51 & $Pmam$ && 7 & $p2mg$ & $22 \mathord{\ast}$ & $-$ && 7 & $(p2mg)$ & $22 \mathord{\ast}$ & $-$ \\
41& $pmma$ & $-$ && $mmm$ & $D_{2h}$ & $\mathord{\ast} 222$ && 51 & $Pmma$ && 6 & $p2mm$ & $\mathord{\ast} 2222$ & $-$ && 6 & $(p2mm)$ & $\mathord{\ast} 2222$ & $-$ \\
42& $pman$ & $-$ && $mmm$ & $D_{2h}$ & $\mathord{\ast} 222$ && 53 & $Pman$ && 9 & $c2mm$ & $2 \mathord{\ast} 22$ & $-$ && 7 & $(p2mg)$ & $22 \mathord{\ast}$ & $-$ \\
43& $pbaa$ & $-$ && $mmm$ & $D_{2h}$ & $\mathord{\ast} 222$ && 54 & $Pbaa$ && 7 & $p2mg$ & $22 \mathord{\ast}$ & $-$ && 8 & $(p2gg)$ & $22 \mathord{\times}$ & $-$ \\
44& $pbam$ & $-$ && $mmm$ & $D_{2h}$ & $\mathord{\ast} 222$ && 55 & $Pbam$ && 8 & $p2gg$ & $22 \mathord{\times}$ & $-$ && 8 & $(p2gg)$ & $22 \mathord{\times}$ & $-$ \\
45& $pbma$ & $-$ && $mmm$ & $D_{2h}$ & $\mathord{\ast} 222$ && 57 & $Pbma$ && 7 & $p2mg$ & $22 \mathord{\ast}$ & $-$ && 7 & $(p2mg)$ & $22 \mathord{\ast}$ & $-$ \\
46& $pmmn$ & $-$ && $mmm$ & $D_{2h}$ & $\mathord{\ast} 222$ && 59 & $Pmmn$ && 9 & $c2mm$ & $2 \mathord{\ast} 22$ & $-$ && 6 & $(p2mm)$ & $\mathord{\ast} 2222$ & $-$ \\
47& $cmmm$ & $-$ && $mmm$ & $D_{2h}$ & $\mathord{\ast} 222$ && 65 & $Cmmm$ && 9 & $c2mm$ & $2 \mathord{\ast} 22$ & $-$ && 9 & $(c2mm)$ & $2 \mathord{\ast} 22$ & $-$ \\
48& $cmme$ & $-$ && $mmm$ & $D_{2h}$ & $\mathord{\ast} 222$ && 67 & $Cmme$ && 6 & $p2mm$ & $\mathord{\ast} 2222$ & $-$ && 9 & $(c2mm)$ & $2 \mathord{\ast} 22$ & $-$ \\
\hline 49& $p4$ & $+$ && $4$ & $C_4$ & $44$ && 75 & $P4$ && 10 & $p4$ & $442$ & $+$ && 10 & $p4$ & $442$ & $+$ \\
50& $p\bar{4}$ & $-$ && $\bar{4}$ & $C_{i4}$ & $2\mathord{\times}$ && 81 & $P\bar{4}$ && 10 & $p4$ & $442$ & $+$ && 2 & $(p2)$ & $2222$ & $+$ \\
51& $p4/m$ & $-$ && $4/m$ & $C_{4h}$ & $4\mathord{\ast}$ && 83 & $P4/m$ && 10 & $p4$ & $442$ & $+$ && 10 & $(p4)$ & $442$ & $+$ \\
52& $p4/n$ & $-$ && $4/m$ & $C_{4h}$ & $4\mathord{\ast}$ && 85 & $P4/n$ && 12 & $p4gm$ & $4 \mathord{\ast} 2$ & $-$ && 10 & $(p4)$ & $442$ & $+$ \\
53& $p422$ & $+$ && $422$ & $D_4$ & $422$ && 89 & $P422$ && 11 & $p4mm$ & $\mathord{\ast} 442$ & $-$ && 10 & $(p4)$ & $442$ & $+$ \\
54& $p42_1 2$ & $+$ && $422$ & $D_4$ & $422$ && 90 & $P42_1 2$ && 12 & $p4gm$ & $4 \mathord{\ast} 2$ & $-$ && 10 & $(p4)$ & $442$ & $+$ \\
55& $p4mm$ & $-$ && $4mm$ & $C_{4v}$ & $\mathord{\ast} 44$ && 99 & $P4mm$ && 11 & $p4mm$ & $\mathord{\ast} 442$ & $-$ && 11 & $p4mm$ & $\mathord{\ast} 442$ & $-$ \\
56& $p4bm$ & $-$ && $4mm$ & $C_{4v}$ & $\mathord{\ast} 44$ && 100 & $P4bm$ && 12 & $p4gm$ & $4 \mathord{\ast} 2$ & $-$ && 12 & $p4gm$ & $4 \mathord{\ast} 2$ & $-$ \\
57& $p\bar{4}2m$ & $-$ && $\bar{4}2m$ & $D_{2d}$ & $2\mathord{\ast} 2$ && 111 & $P\bar{4}2m$ && 11 & $p4mm$ & $\mathord{\ast} 442$ & $-$ && 9 & $(c2mm)$ & $2 \mathord{\ast} 22$ & $-$ \\
58& $p\bar{4}2_1 m$ & $-$ && $\bar{4}2m$ & $D_{2d}$ & $2\mathord{\ast} 2$ && 113 & $P\bar{4}2_1 m$ && 12 & $p4gm$ & $4 \mathord{\ast} 2$ & $-$ && 9 & $(c2mm)$ & $2 \mathord{\ast} 22$ & $-$ \\
59& $p\bar{4}m2$ & $-$ && $(\bar{4}m2)$ & $D_{2d}$ & $2\mathord{\ast} 2$ && 115 & $P\bar{4}m2$ && 11 & $p4mm$ & $\mathord{\ast} 442$ & $-$ && 6 & $(p2mm)$ & $\mathord{\ast} 2222$ & $-$ \\
60& $p\bar{4}b2$ & $-$ && $(\bar{4}m2)$ & $D_{2d}$ & $2\mathord{\ast} 2$ && 117 & $P\bar{4}b2$ && 12 & $p4gm$ & $4 \mathord{\ast} 2$ & $-$ && 8 & $(p2gg)$ & $22 \mathord{\times}$ & $-$ \\
61& $p4/mmm$ & $-$ && $4/mmm$ & $D_{4h}$ & $\mathord{\ast} 422$ && 123 & $P4/mmm$ && 11 & $p4mm$ & $\mathord{\ast} 442$ & $-$ && 11 & $(p4mm)$ & $\mathord{\ast} 442$ & $-$ \\
62& $p4/nbm$ & $-$ && $4/mmm$ & $D_{4h}$ & $\mathord{\ast} 422$ && 125 & $P4/nbm$ && 11 & $p4mm$ & $\mathord{\ast} 442$ & $-$ && 12 & $(p4gm)$ & $4 \mathord{\ast} 2$ & $-$ \\
63& $p4/mbm$ & $-$ && $4/mmm$ & $D_{4h}$ & $\mathord{\ast} 422$ && 127 & $P4/mbm$ && 12 & $p4gm$ & $4 \mathord{\ast} 2$ & $-$ && 12 & $(p4gm)$ & $4 \mathord{\ast} 2$ & $-$ \\
64& $p4/nmm$ & $-$ && $4/mmm$ & $D_{4h}$ & $\mathord{\ast} 422$ && 129 & $P4/nmm$ && 11 & $p4mm$ & $\mathord{\ast} 442$ & $-$ && 11 & $(p4mm)$ & $\mathord{\ast} 442$ & $-$ \\
\hline 65& $p3$ & $+$ && $3$ & $C_3$ & $33$ && 143 & $P3$ && 13 & $p3$ & $333$ & $+$ && 13 & $p3$ & $333$ & $+$ \\
66& $p\bar{3}$ & $-$ && $\bar{3}$ & $C_{i3}$ & $3\mathord{\times}$ && 147 & $P\bar{3}$ && 16 & $p6$ & $632$ & $+$ && 13 & $(p3)$ & $333$ & $+$ \\
67& $p312$ & $+$ && $312$ & $D_3$ & $322$ && 149 & $P312$ && 14 & $p3m1$ & $\mathord{\ast} 333$ & $-$ && 13 & $(p3)$ & $333$ & $+$ \\
68& $p321$ & $+$ && $(321)$ & $D_3$ & $322$ && 150 & $P321$ && 15 & $p31m$ & $3 \mathord{\ast} 3$ & $-$ && 13 & $(p3)$ & $333$ & $+$ \\
69& $p3m1$ & $-$ && $3m1$ & $C_{3v}$ & $\mathord{\ast} 33$ && 156 & $P3m1$ && 14 & $p3m1$ & $\mathord{\ast} 333$ & $-$ && 14 & $p3m1$ & $\mathord{\ast} 333$ & $-$ \\
70& $p31m$ & $-$ && $(31m)$ & $C_{3v}$ & $\mathord{\ast} 33$ && 157 & $P31m$ && 15 & $p31m$ & $3 \mathord{\ast} 3$ & $-$ && 15 & $p31m$ & $3 \mathord{\ast} 3$ & $-$ \\
71& $p\bar{3}1m$ & $-$ && $\bar{3}1m$ & $T_h$ & $3 \mathord{\ast} 2$ && 162 & $P\bar{3}1m$ && 17 & $p6mm$ & $\mathord{\ast} 632$ & $-$ && 15 & $(p31m)$ & $3 \mathord{\ast} 3$ & $-$ \\
72& $p\bar{3}m1$ & $-$ && $(\bar{3}m1)$ & $T_h$ & $3 \mathord{\ast} 2$ && 164 & $P\bar{3}m1$ && 17 & $p6mm$ & $\mathord{\ast} 632$ & $-$ && 14 & $(p3m1)$ & $\mathord{\ast} 333$ & $-$ \\
\hline 73& $p6$ & $+$ && $6$ & $C_6$ & $66$ && 168 & $P6$ && 16 & $p6$ & $632$ & $+$ && 16 & $p6$ & $632$ & $+$ \\
74& $p\bar{6}$ & $-$ && $\bar{6}$ & $C_{3h}$ & $3 \mathord{\ast}$ && 174 & $P\bar{6}$ && 13 & $p3$ & $333$ & $+$ && 13 & $(p3)$ & $333$ & $+$ \\
75& $p6/m$ & $-$ && $6/m$ & $C_{6h}$ & $6 \mathord{\ast}$ && 175 & $P6/m$ && 16 & $p6$ & $632$ & $+$ && 16 & $(p6)$ & $632$ & $+$ \\
76& $p622$ & $+$ && $622$ & $D_6$ & $622$ && 177 & $P622$ && 17 & $p6mm$ & $\mathord{\ast} 632$ & $-$ && 16 & $(p6)$ & $632$ & $+$ \\
77& $p6mm$ & $-$ && $6mm$ & $C_{6v}$ & $\mathord{\ast} 66$ && 183 & $P6mm$ && 17 & $p6mm$ & $\mathord{\ast} 632$ & $-$ && 17 & $p6mm$ & $\mathord{\ast} 632$ & $-$ \\
78& $p\bar{6}m2$ & $-$ && $\bar{6}m2$ & $D_{3h}$ & $\mathord{\ast} 322$ && 187 & $P\bar{6}m2$ && 14 & $p3m1$ & $\mathord{\ast} 333$ & $-$ && 14 & $(p3m1)$ & $\mathord{\ast} 333$ & $-$ \\
79& $p\bar{6}2m$ & $-$ && $(\bar{6}2m)$ & $D_{3h}$ & $\ast 322$ && 189 & $P\bar{6}2m$ && 15 & $p31m$ & $3 \mathord{\ast} 3$ & $-$ && 15 & $(p31m)$ & $3 \mathord{\ast} 3$ & $-$ \\
80& $p6/mmm$ & $-$ && $6/mmm$ & $D_{6h}$ & $\mathord{\ast} 622$ && 191 & $P6/mmm$ && 17 & $p6mm$ & $\mathord{\ast} 632$ & $-$ && 17 & $(p6mm)$ & $\mathord{\ast} 632$ & $-$ \\
\end{longtable}
\normalsize

%%%%%%%%%%% CAPTIONS FOR OTHER SUPPLEMENTARY FILES %%%%%%%%%%

%\clearpage % Clear all remaining figures and tables then start a new page

%\paragraph{Caption for Movie S1.}
%\textbf{All captions must start with a short bold sentence, acting as a title.}
%Then explain what is shown in the supplementary video file.
%Give as much detail as you would for a figure e.g. explain axes, color maps etc.
%If the video is an animated equivalent of one of the static figures, state e.g.
%`Animated version of Figure~\ref{fig:example}.'

%\paragraph{Caption for Data S1.}
%\textbf{All captions must start with a short bold sentence, acting as a title.}
%Then explain what is included in the supplementary data file.
%Give as much detail as you would for a table e.g. explain the meaning of every column,
%units used, any special notation etc.

%%%%%%%%%%%%%%%% SUPPLEMENTARY REFERENCES %%%%%%%%%%%%%%%

% Do NOT include a reference list in the supplement.
% All references must be in a single list at the end of the main text.
% The copyeditors will ensure that the correct reference list appears with each version of the paper
% (print, HTML, PDF, mobile app, metadata for bibliographic databases etc.)


\begin{thebibliography}{10}
\providecommand{\url}[1]{\texttt{#1}}
\expandafter\ifx\csname urlstyle\endcsname\relax
  \providecommand{\doi}[1]{doi:\discretionary{}{}{}#1}\else
  \providecommand{\doi}{doi:\discretionary{}{}{}\begingroup \urlstyle{rm}\Url}\fi

\bibitem{Herrmann1929zkrist}
K.~Herrmann, E.~Alexander, {XIII}. {D}ie 80 zweidimensionalen {R}aumgruppen. \emph{Zeitschrift f\"{u}r Kristallographie - Crystalline Materials} \textbf{70}~(1--6), 328--345 (1929), \url{\url{https://doi.org/10.1524/zkri.1929.70.1.328}}.

\bibitem{Weber1929zkristallogr}
L.~Weber, {XII}. Die {S}ymmetrie homogener ebener {P}unktsysteme. \emph{Z. Kristallogr. Cryst. Mater.} \textbf{70}, 309--327 (1929), \doi{10.1524/zkri.1929.70.1.309}, \url{\url{https://doi.org/10.1524/zkri.1929.70.1.309}}.

\bibitem{IT_E2006}
V.~Kopsk\'{y}, D.~B. Litvin, \emph{Subperiodic groups}, vol.~E of \emph{International Tables for Crystallography} (2006), \doi{10.1107/97809553602060000647}, \url{\url{https://doi.org/10.1107/97809553602060000647}}.

\bibitem{Markande2020bridges}
S.~G. Markande, E.~A. Matsumoto, Knotty Knits are Tangles on Tori, in \emph{Proceedings of Bridges 2020: Mathematics, Art, Music, Architecture, Education, Culture} (Tessellations Publishing, Phoenix, Arizona) (2020), pp. 103--112, \url{\url{http://archive.bridgesmathart.org/2020/bridges2020-103.html}}.

\bibitem{Diamantis2024Symmetry}
I.~Diamantis, S.~Lambropoulou, S.~Mahmoudi, Directional Invariants of Doubly Periodic Tangles. \emph{Symmetry} \textbf{16}, 968 (2024), \doi{10.3390/sym16080968}, \url{\url{https://doi.org/10.3390/sym16080968}}.

\bibitem{DeLasPenas2024acta}
M.~L.~A. De~Las Pe\~{n}as, M.~Tomenes, K.~Liza, Symmetry groups of two-way twofold and three-way threefold fabrics. \emph{Acta Crystallographica} \textbf{A80}, 33--51 (2024), \url{\url{https://doi.org/10.1107/S2053273323008938}}.

\bibitem{Sannai1998tower}
Y.~Okada, \emph{et~al.}, {S}annai {M}aruyama {A}rchaeological {S}ite {IX}, {S}ith {T}ower {A}rea Investigation Report, in \emph{Aomori Prefecture Buried Cultural Properties Investigation Report}, {Culture Division, Education Bureau, Aomori Prefectural Government}, Ed. (Aomori Prefectural Board of Education), vol. 249 (1998), \url{\url{https://doi.org/10.24484/sitereports.34142}}.

\bibitem{Noshiro2019archaeol}
S.~Noshiro, Y.~Sasaki, K.~Kobayashia, M.~Suzuki, I.~Nishida, Material selection and weaving techniques for the oldest basketry in {J}apan found at the {H}igashimyou site, {S}aga {P}refecture. \emph{Journal of Archaeological Science: Reports} \textbf{23}, 12--24 (2019), \url{\url{https://doi.org/10.1016/j.jasrep.2018.10.009}}.

\bibitem{Pfister1945dura}
R.~Pfister, L.~Bellinger, The textiles, in \emph{Excavations at {D}ura-{E}uropos} (Yale University Press, New Haven), vol. Final report IV, part II (1945), \url{\url{}}.

\bibitem{Grishanov2009textres}
S.~Grishanov, V.~Meshkov, A.~Omelchenko, A Topological Study of Textile Structures. Part I: An Introduction to Topological Methods. \emph{Textile Research Journal} \textbf{79}, 702--713 (2009), \url{\url{https://doi.org/10.1177/0040517508095600}}.

\bibitem{Kauffman2006formal}
L.~H. Kauffman, \emph{Formal Knot Theory}, Dover Books on Mathematics (Dover Publications) (2006), \url{\url{}}.

\bibitem{Purcell2020Hyperbolic}
J.~S. Purcell, \emph{Hyperbolic Knot Theory}, no. 209 in Graduate Studies in Mathematics (2020).

\bibitem{Kauffman1990transamermathsoc}
L.~H. Kauffman, An Invariant of Regular Isotopy. \emph{Transactions of the American Mathematical Society} \textbf{318}~(2), 417--471 (1990), \url{\url{https://doi.org/10.1090/S0002-9947-1990-0958895-7}}.

\bibitem{Patil2020sci}
V.~P. Patil, J.~D. Sandt, M.~Kolle, J.~Dunkel, Topological mechanics of knots and tangles. \emph{Science} \textbf{367}, 71--75 (2020), \url{\url{https://doi.org/10.1126/science.aaz0135}}.

\bibitem{WIREROPEUSERSMANUAL2nd1981}
{COMMITTEE OF WIRE ROPE PRODUCERS American Iron}, {Steel Institute and THE WIRE ROPE TECHNICAL BOARD}, \emph{WIRE ROPE USERS MANUAL}, 2nd ed. (1981).

\bibitem{deGennes1979polymer}
P.-G. de~Gennes, \emph{Scaling Concepts in Polymer Physics} (Cornell University Press) (1979), \url{\url{}}.

\bibitem{Miura2020form}
K.~Miura, S.~Pellegrino, \emph{Forms and Concepts for Lightweight Structures} (Cambridge University Press) (2020), \url{\url{https://doi.org/10.1017/9781139048569}}.

\bibitem{Flaum2024sci}
E.~Flaum, M.~Prakash, Curved crease origami and topological singularities enable hyperextensibility of \textit{L. olor}. \emph{Science} \textbf{384}, eadk5511 (2024), \url{\url{https://doi.org/10.1126/science.adk5511}}.

\bibitem{Singal2024natcommun}
K.~Singal, \emph{et~al.}, Programming mechanics in knitted materials, stitch by stitch. \emph{Nature Communications} \textbf{15}, 2622 (2024), \url{\url{https://doi.org/10.1038/s41467-024-46498-z}}.

\bibitem{supplmater}
Supplementary Materials: materials and methods, additional figures and tables; 3D models are available via Zenodo~\cite{repository}.

\bibitem{Barron2012chirality}
L.~D. Barron, From Cosmic Chirality to Protein Structure: {L}ord {K}elvin's Legacy. \emph{Chirality} \textbf{24}, 879--893 (2012), \doi{10.1002/chir.22017}, \url{\url{https://doi.org/10.1002/chir.22017}}.

\bibitem{Penrose1958psy}
L.~Penrose, R.~Penrose, IMPOSSIBLE OBJECTS: A SPECIAL TYPE OF VISUAL ILLUSION. \emph{British Journal of Psychology} \textbf{49}, 31--33 (1958), \url{\url{https://doi.org/10.1111/j.2044-8295.1958.tb00634.x}}.

\bibitem{Jakli2013liqcrystrev}
A.~J\'{a}kli, Liquid crystals of the twenty-first century – nematic phase of bent-core molecules. \emph{Liquid Crystals Reviews} \textbf{1}, 65--82 (2013), \url{\url{https://doi.org/10.1080/21680396.2013.803701}}.

\bibitem{Arago1811}
D.~F.~J. Arago, M\'{e}moire sur une modification remarquable qu'\'{e}prouvent les rayons lumineux dans leur passage \`{a} travers certains corps diaphanes et sur quelques autres nouveaux ph\'{e}nom\`{e}nes d'optique. \emph{M\'{e}moires de la classe des sciences math\'{e}matiques et physiques de l'Institut Imp\'{e}rial de France} \textbf{12}, 93--134 (1811), \url{\url{}}.

\bibitem{Nespolo2021japplcrystallogr}
M.~Nespolo, A.~H. Benahsene, Symmetry and chirality in crystals. \emph{J. Appl. Crystallogr.} \textbf{54}, 1594--1599 (2021), \url{\url{https://doi.org/10.1107/S1600576721009109}}.

\bibitem{Rekis2020ActaCrystB}
T.~Rekis, Crystallization of chiral molecular compounds: what can be learned from the Cambridge Structural Database? \emph{Acta Cryst. B} \textbf{76}, 307--315 (2020), \doi{10.1107/S2052520620003601}, \url{\url{https://doi.org/10.1107/S2052520620003601}}.

\bibitem{Liu2018chemsocrev}
Y.~Liu, M.~O’Keeﬀe, M.~M.~J. Treacy, O.~M. Yaghi, The geometry of periodic knots, polycatenanes and weaving from a chemical perspective: a library for reticular chemistry. \emph{Chem. Soc. Rev.} \textbf{47}, 4642 (2018), \url{\url{https://doi.org/10.1039/c7cs00695k}}.

\bibitem{Klotz2024soft}
A.~R. Klotz, C.~J. Andersona, M.~S. Dimitriyev, Chirality effects in molecular chainmail. \emph{Soft Matter} \textbf{20}, 7044 (2024), \url{\url{https://doi.org/10.1039/D4SM00729H}}.

\bibitem{Wijnhoven2021}
M.~A. Wijnhoven, \emph{European Mail Armour: Ringed Battle Shirts from the Iron Age, Roman Period and Early Middle Ages} (Amsterdam University Press) (2021), \doi{10.1017/9789048554294}, \url{\url{https://doi.org/10.1017/9789048554294}}.

\bibitem{Wetzel2024bridges}
D.~Wetzel, P.~Gailiunas, M.~Gaither-Ganim, W.~Holt, Triply Periodic Helical Weaves, in \emph{Proceedings of Bridges 2024: Mathematics, Art, Music, Architecture, Culture}, H.~Verrill, K.~Kattchee, S.~L. Gould, E.~Torrence, Eds. (Tessellations Publishing, Phoenix, Arizona) (2024), pp. 267--274, \url{\url{http://archive.bridgesmathart.org/2024/bridges2024-267.html}}.

\bibitem{repository}
Repository: 3{D} models of fabrics available at {Z}enodo (temporarily restricted to co-authors, publicly accessible upon publication), \doi{10.5281/zenodo.17156949}.

\bibitem{Langer1996siamreview}
J.~Langer, D.~Singer, {L}agrangian Aspects of the {K}irchhoff Elastic Rod. \emph{SIAM Review} \textbf{38}, 605--618 (1996), \url{\url{https://www.jstor.org/stable/2132934}}.

\bibitem{Matsumoto2025jacs}
A.~Matsumoto, \emph{et~al.}, How Temperature Change Affects the Lattice Parameters, Molecular Conformation, and Reaction Cavity in Enantiomeric and Racemic Crystals of Thalidomide. \emph{J. Am. Chem. Soc.} \textbf{147}, 11988--11997 (2025), \url{\url{https://doi.org/10.1021/jacs.4c18394}}.

\bibitem{Togawa2012prl}
Y.~Togawa, \emph{et~al.}, Chiral Magnetic Soliton Lattice on a Chiral Helimagnet. \emph{Phys. Rev. Lett.} \textbf{108}, 107202 (2012), \url{\url{https://doi.org/10.1103/PhysRevLett.108.107202}}.

\bibitem{Frank2020bridges}
F.~A. Farris, Wallpaper Patterns from Nonplanar Chain Mail Links, in \emph{Proceedings of Bridges 2020: Mathematics, Art, Music, Architecture, Education, Culture}, C.~Yackel, R.~Bosch, E.~Torrence, K.~Fenyvesi, Eds. (Tessellations Publishing, Phoenix, Arizona) (2020), pp. 183--190, \url{\url{http://archive.bridgesmathart.org/2020/bridges2020-183.html}}.

\bibitem{IntTabCryst2016A}
M.~I. Aroyo, ed., \emph{SPACE-GROUP SYMMETRY}, vol.~A of \emph{INTERNATIONAL TABLES FOR CRYSTALLOGRAPHY} (Wiley), 6 ed. (2016), \url{\url{}}.

\bibitem{Conway2008symmetry}
J.~H. Conway, H.~Burgiel, C.~Goodman-Strauss, \emph{The Symmetries of Things} (CRC Press \& Taylor \& Francis Group) (2008).

\end{thebibliography}
\end{document}